\documentclass[longauth]{aa}

\usepackage{graphicx}
\usepackage[dvipsnames]{xcolor}
\usepackage[T1]{fontenc}
\usepackage[utf8]{inputenc}
\usepackage{lmodern}
\usepackage{txfonts}
\usepackage{multirow}

\usepackage{euclid}
\usepackage{bm}
\usepackage[normalem]{ulem}

\usepackage[colorlinks,bookmarks]{hyperref}
\definecolor{linkblue}{rgb}{0,0,0.8}
\definecolor{linkgreen}{rgb}{0,0.5,0}
\hypersetup{linkcolor=linkblue, citecolor=linkgreen, urlcolor=linkblue}

\usepackage[nameinlink,noabbrev]{cleveref}
\Crefname{equation}{Eq.}{Eqs.}
\Crefname{eqnarray}{Eq.}{Eqs.}
\Crefname{section}{Sect.}{Sects.}
\Crefname{figure}{Fig.}{Figs.}
\crefname{equation}{Equation}{Equations}
\crefname{section}{Sect.}{Sect.}
\crefname{figure}{Figure}{Figures}

\newcommand{\aperp}{a_{\perp }}
\newcommand{\aperpo}{a_{\perp 0}}
\newcommand{\adotperp}{\dot{a}_{\perp}}
\newcommand{\apar}{a_{\parallel}}
\newcommand{\adotpar}{\dot{a}_{\parallel}}

\newcommand{\Hperp}{H_{\perp}}
\newcommand{\Hperpo}{H_{\perp 0}}
\newcommand{\Hpar}{H_{\parallel}}

\def\be{\begin{equation}}
\def\ee{\end{equation}}
\def\ba{\begin{eqnarray}}
\def\ea{\end{eqnarray}}
\def\eqi{\begin{equation}}
\def\eqf{\end{equation}}
\def\eqia{\begin{eqnarray}}
\def\eqfa{\end{eqnarray}}
\def\lcdm{$\Lambda$CDM }
\def\lltb{$\Lambda$LTB }

\def\d {\mathrm{d}}

\newcommand{\orcid}[1]{} 
\defcitealias{IST:paper1}{EC20}
\graphicspath{{./}}

\begin{document} 

\title{\Euclid: Testing the Copernican principle with next-generation surveys\thanks{This paper is published on behalf of the Euclid Consortium.}}

\titlerunning{\Euclid: Testing the Copernican principle with next-generation surveys}


\author{D.~Camarena\orcid{0000-0001-7165-0439}$^{1}$\thanks{\email{dacato115@gmail.com}}, V.~Marra\orcid{0000-0002-7773-1579}$^{2,3,4}$, Z.~Sakr\orcid{0000-0002-4823-3757}$^{5,6,7}$, S.~Nesseris\orcid{0000-0002-0567-0324}$^{8}$, A.~Da Silva\orcid{0000-0002-6385-1609}$^{9,10}$, J.~Garcia-Bellido\orcid{0000-0002-9370-8360}$^{8}$, P.~Fleury\orcid{0000-0001-9292-3651}$^{11}$, L.~Lombriser$^{12}$, M.~Martinelli\orcid{0000-0002-6943-7732}$^{13}$, C.~J.~A.~P.~Martins\orcid{0000-0002-4886-9261}$^{14,15}$
% ...
}

\institute{$^{1}$ PPGCosmo, Universidade Federal do Esp\'{i}rito Santo, 29075-910,Vit\'{o}ria, ES, Brazil\\
$^{2}$ N\'{u}cleo Cosmo-ufes \& Departamento de F\'{i}sica, Universidade Federal do Esp\'{i}rito Santo, 29075-910, Vit\'{o}ria, ES, Brazil\\
$^{3}$ INAF-Osservatorio Astronomico di Trieste, Via G. B. Tiepolo 11, I-34143 Trieste, Italy\\
$^{4}$ IFPU, Institute for Fundamental Physics of the Universe, via Beirut 2, 34151 Trieste, Italy\\
$^{5}$ Institut de Recherche en Astrophysique et Plan\'etologie (IRAP), Universit\'e de Toulouse, CNRS, UPS, CNES, 14 Av. Edouard Belin, F-31400 Toulouse, France\\
$^{6}$ Institut f\"ur Theoretische Physik, University of Heidelberg, Philosophenweg 16, 69120 Heidelberg, Germany\\
$^{7}$ Universit\'e St Joseph; Faculty of Sciences, Beirut, Lebanon\\
$^{8}$ Instituto de F\'isica Te\'orica UAM-CSIC, Campus de Cantoblanco, E-28049 Madrid, Spain\\
$^{9}$ Departamento de F\'isica, Faculdade de Ci\^encias, Universidade de Lisboa, Edif\'icio C8, Campo Grande, PT1749-016 Lisboa, Portugal\\
$^{10}$ Instituto de Astrof\'isica e Ci\^encias do Espa\c{c}o, Faculdade de Ci\^encias, Universidade de Lisboa, Campo Grande, PT-1749-016 Lisboa, Portugal\\
$^{11}$ Institut de Physique Th\'eorique, CEA, CNRS, Universit\'e Paris-Saclay F-91191 Gif-sur-Yvette Cedex, France\\
$^{12}$ Universit\'e de Gen\`eve, D\'epartement de Physique Th\'eorique and Centre for Astroparticle Physics, 24 quai Ernest-Ansermet, CH-1211 Gen\`eve 4, Switzerland\\
$^{13}$ INAF-Osservatorio Astronomico di Roma, Via Frascati 33, I-00078 Monteporzio Catone, Italy\\
$^{14}$ Centro de Astrof\'{\i}sica da Universidade do Porto, Rua das Estrelas, 4150-762 Porto, Portugal\\
$^{15}$ Instituto de Astrof\'isica e Ci\^encias do Espa\c{c}o, Universidade do Porto, CAUP, Rua das Estrelas, PT4150-762 Porto, Portugal\\
$^{16}$ Departamento de F\'isica, FCFM, Universidad de Chile, Blanco Encalada 2008, Santiago, Chile\\
$^{17}$ School of Physics and Astronomy, Queen Mary University of London, Mile End Road, London E1 4NS, UK\\
$^{18}$ Dipartimento di Fisica, Universit\'a degli Studi di Torino, Via P. Giuria 1, I-10125 Torino, Italy\\
$^{19}$ INFN-Sezione di Torino, Via P. Giuria 1, I-10125 Torino, Italy\\
$^{20}$ INAF-Osservatorio Astrofisico di Torino, Via Osservatorio 20, I-10025 Pino Torinese (TO), Italy\\
$^{21}$ INAF-IASF Milano, Via Alfonso Corti 12, I-20133 Milano, Italy\\
$^{22}$ Institute for Theoretical Particle Physics and Cosmology (TTK), RWTH Aachen University, D-52056 Aachen, Germany\\
$^{23}$ Universit\'{e} Paris-Saclay, CNRS/IN2P3, IJCLab, 91405 Orsay, France\\
$^{24}$ Centre National d'Etudes Spatiales, Toulouse, France\\
$^{25}$ Universit\'e Paris-Saclay, Universit\'e Paris Cit\'e, CEA, CNRS, Astrophysique, Instrumentation et Mod\'elisation Paris-Saclay, 91191 Gif-sur-Yvette, France\\
$^{26}$ Institute of Space Sciences (ICE, CSIC), Campus UAB, Carrer de Can Magrans, s/n, 08193 Barcelona, Spain\\
$^{27}$ Institut d'Estudis Espacials de Catalunya (IEEC), Carrer Gran Capit\'a 2-4, 08034 Barcelona, Spain\\
$^{28}$ Universit\'e Paris-Saclay, CNRS, Institut d'astrophysique spatiale, 91405, Orsay, France\\
$^{29}$ ESAC/ESA, Camino Bajo del Castillo, s/n., Urb. Villafranca del Castillo, 28692 Villanueva de la Ca\~nada, Madrid, Spain\\
$^{30}$ Institute of Cosmology and Gravitation, University of Portsmouth, Portsmouth PO1 3FX, UK\\
$^{31}$ INAF-Osservatorio di Astrofisica e Scienza dello Spazio di Bologna, Via Piero Gobetti 93/3, I-40129 Bologna, Italy\\
$^{32}$ Dipartimento di Fisica e Astronomia "Augusto Righi" - Alma Mater Studiorum Universit\`{a} di Bologna, via Piero Gobetti 93/2, I-40129 Bologna, Italy\\
$^{33}$ INFN-Sezione di Bologna, Viale Berti Pichat 6/2, I-40127 Bologna, Italy\\
$^{34}$ Dipartimento di Fisica, Universit\'a degli studi di Genova, and INFN-Sezione di Genova, via Dodecaneso 33, I-16146, Genova, Italy\\
$^{35}$ INFN-Sezione di Roma Tre, Via della Vasca Navale 84, I-00146, Roma, Italy\\
$^{36}$ INAF-Osservatorio Astronomico di Capodimonte, Via Moiariello 16, I-80131 Napoli, Italy\\
$^{37}$ Mullard Space Science Laboratory, University College London, Holmbury St Mary, Dorking, Surrey RH5 6NT, UK\\
$^{38}$ Institut de F\'{i}sica d'Altes Energies (IFAE), The Barcelona Institute of Science and Technology, Campus UAB, 08193 Bellaterra (Barcelona), Spain\\
$^{39}$ Port d'Informaci\'{o} Cient\'{i}fica, Campus UAB, C. Albareda s/n, 08193 Bellaterra (Barcelona), Spain\\
$^{40}$ INFN section of Naples, Via Cinthia 6, I-80126, Napoli, Italy\\
$^{41}$ Department of Physics "E. Pancini", University Federico II, Via Cinthia 6, I-80126, Napoli, Italy\\
$^{42}$ Dipartimento di Fisica e Astronomia "Augusto Righi" - Alma Mater Studiorum Universit\'a di Bologna, Viale Berti Pichat 6/2, I-40127 Bologna, Italy\\
$^{43}$ INAF-Osservatorio Astrofisico di Arcetri, Largo E. Fermi 5, I-50125, Firenze, Italy\\
$^{44}$ Institut national de physique nucl\'eaire et de physique des particules, 3 rue Michel-Ange, 75794 Paris C\'edex 16, France\\
$^{45}$ Institute for Astronomy, University of Edinburgh, Royal Observatory, Blackford Hill, Edinburgh EH9 3HJ, UK\\
$^{46}$ European Space Agency/ESRIN, Largo Galileo Galilei 1, 00044 Frascati, Roma, Italy\\
$^{47}$ Univ Lyon, Univ Claude Bernard Lyon 1, CNRS/IN2P3, IP2I Lyon, UMR 5822, F-69622, Villeurbanne, France\\
$^{48}$ Observatoire de Sauverny, Ecole Polytechnique F\'ed\'erale de Lau- sanne, CH-1290 Versoix, Switzerland\\
$^{49}$ Department of Astronomy, University of Geneva, ch. d\'Ecogia 16, CH-1290 Versoix, Switzerland\\
$^{50}$ Department of Physics, Oxford University, Keble Road, Oxford OX1 3RH, UK\\
$^{51}$ Jodrell Bank Centre for Astrophysics, Department of Physics and Astronomy, University of Manchester, Oxford Road, Manchester M13 9PL, UK\\
$^{52}$ INFN-Padova, Via Marzolo 8, I-35131 Padova, Italy\\
$^{53}$ Istituto Nazionale di Astrofisica (INAF) - Osservatorio di Astrofisica e Scienza dello Spazio (OAS), Via Gobetti 93/3, I-40127 Bologna, Italy\\
$^{54}$ Istituto Nazionale di Fisica Nucleare, Sezione di Bologna, Via Irnerio 46, I-40126 Bologna, Italy\\
$^{55}$ INAF-Osservatorio Astronomico di Padova, Via dell'Osservatorio 5, I-35122 Padova, Italy\\
$^{56}$ Max Planck Institute for Extraterrestrial Physics, Giessenbachstr. 1, D-85748 Garching, Germany\\
$^{57}$ Universit\"ats-Sternwarte M\"unchen, Fakult\"at f\"ur Physik, Ludwig-Maximilians-Universit\"at M\"unchen, Scheinerstrasse 1, 81679 M\"unchen, Germany\\
$^{58}$ Institute of Theoretical Astrophysics, University of Oslo, P.O. Box 1029 Blindern, N-0315 Oslo, Norway\\
$^{59}$ Jet Propulsion Laboratory, California Institute of Technology, 4800 Oak Grove Drive, Pasadena, CA, 91109, USA\\
$^{60}$ von Hoerner \& Sulger GmbH, Schlo{\ss}Platz 8, D-68723 Schwetzingen, Germany\\
$^{61}$ Technical University of Denmark, Elektrovej 327, 2800 Kgs. Lyngby, Denmark\\
$^{62}$ Max-Planck-Institut f\"ur Astronomie, K\"onigstuhl 17, D-69117 Heidelberg, Germany\\
$^{63}$ Department of Physics and Helsinki Institute of Physics, Gustaf H\"allstr\"omin katu 2, 00014 University of Helsinki, Finland\\
$^{64}$ NOVA optical infrared instrumentation group at ASTRON, Oude Hoogeveensedijk 4, 7991PD, Dwingeloo, The Netherlands\\
$^{65}$ Argelander-Institut f\"ur Astronomie, Universit\"at Bonn, Auf dem H\"ugel 71, 53121 Bonn, Germany\\
$^{66}$ Department of Physics, Institute for Computational Cosmology, Durham University, South Road, DH1 3LE, UK\\
$^{67}$ INFN-Bologna, Via Irnerio 46, I-40126 Bologna, Italy\\
$^{68}$ Institute of Physics, Laboratory of Astrophysics, Ecole Polytechnique F\'{e}d\'{e}rale de Lausanne (EPFL), Observatoire de Sauverny, 1290 Versoix, Switzerland\\
$^{69}$ European Space Agency/ESTEC, Keplerlaan 1, 2201 AZ Noordwijk, The Netherlands\\
$^{70}$ Department of Physics and Astronomy, University of Aarhus, Ny Munkegade 120, DK-8000 Aarhus C, Denmark\\
$^{71}$ Space Science Data Center, Italian Space Agency, via del Politecnico snc, 00133 Roma, Italy\\
$^{72}$ Institute of Space Science, Bucharest, Ro-077125, Romania\\
$^{73}$ Instituto de Astrof\'isica de Canarias, Calle V\'ia L\'actea s/n, E-38204, San Crist\'obal de La Laguna, Tenerife, Spain\\
$^{74}$ Departamento de Astrof\'{i}sica, Universidad de La Laguna, E-38206, La Laguna, Tenerife, Spain\\
$^{75}$ Aix-Marseille Univ, CNRS/IN2P3, CPPM, Marseille, France\\
$^{76}$ Dipartimento di Fisica e Astronomia "G.Galilei", Universit\'a di Padova, Via Marzolo 8, I-35131 Padova, Italy\\
$^{77}$ Aix-Marseille Univ, CNRS, CNES, LAM, Marseille, France\\
$^{78}$ Centro de Investigaciones Energ\'eticas, Medioambientales y Tecnol\'ogicas (CIEMAT), Avenida Complutense 40, 28040 Madrid, Spain\\
$^{79}$ Instituto de Astrof\'isica e Ci\^encias do Espa\c{c}o, Faculdade de Ci\^encias, Universidade de Lisboa, Tapada da Ajuda, PT-1349-018 Lisboa, Portugal\\
$^{80}$ Universidad Polit\'ecnica de Cartagena, Departamento de Electr\'onica y Tecnolog\'ia de Computadoras, 30202 Cartagena, Spain\\
$^{81}$ Kapteyn Astronomical Institute, University of Groningen, PO Box 800, 9700 AV Groningen, The Netherlands\\
$^{82}$ Infrared Processing and Analysis Center, California Institute of Technology, Pasadena, CA 91125, USA\\
$^{83}$ INAF-Osservatorio Astronomico di Brera, Via Brera 28, I-20122 Milano, Italy\\
$^{84}$ Institut d'Astrophysique de Paris, UMR 7095, CNRS, and Sorbonne Universit\'e, 98 bis boulevard Arago, 75014 Paris, France\\
$^{85}$ Junia, EPA department, F 59000 Lille, France}

\authorrunning{D.~Camarena et al.}

\abstract
%
%
{The Copernican principle, the notion that we are not at a special location in the Universe, is one of the cornerstones of modern cosmology. Its violation would invalidate the Friedmann-Lema\^{\i}tre-Robertson-Walker metric, causing a major change in our understanding of the Universe. Thus, it is of fundamental importance to perform observational tests of this principle.
}
{We determine the precision with which future surveys will be able to test the Copernican principle and their ability to detect any possible violations.
}
{We forecast constraints on the inhomogeneous Lema\^{\i}tre-Tolman-Bondi (LTB) model with a cosmological constant $\Lambda$, basically a cosmological constant $\Lambda$ and cold dark matter (CDM) model but endowed with a spherical inhomogeneity. We consider combinations of currently available data and simulated \Euclid  data, together with external data products, based on both \lcdm and \lltb fiducial models. These constraints are compared to the expectations from the Copernican principle.
}
{When considering the \lcdm fiducial model, we find that \Euclid data, in combination with other current and forthcoming surveys, will improve the constraints on the Copernican principle by about $30\%$, with $\pm10\%$ variations depending on the observables and scales considered.
On the other hand, when considering a \lltb fiducial model, we find that future \Euclid data, combined with other current and forthcoming datasets, will be able to detect gigaparsec-scale inhomogeneities of contrast $-0.1$.
}
{Next-generation surveys, such as \Euclid, will thoroughly test homogeneity at large scales, tightening the constraints on possible violations of the Copernican principle.
}


\maketitle
%
\section{Introduction} \label{sec:intro}

Modern cosmology relies on several fundamental assumptions, such as the hypothesis that we do not occupy a special location in the Universe. Thanks to this assumption, called the Copernican principle, cosmologists have made tremendous advances in the understanding of the Universe and laid the foundation of the standard cosmological model. The Copernican principle, along with the fact that the Universe appears to be statistically isotropic, implies that our Universe is homogeneous and isotropic on sufficiently large scales, eliminating any possible spatial dependence in the cosmological parameters. Equivalently, the space-time is accurately described by the Friedmann-Lema\^{\i}tre-Robertson-Walker (FLRW) metric. Clearly, any violation of the Copernican principle indicates a breakdown of the FLRW paradigm and, therefore, of the standard cosmological model. Thus, testing the Copernican principle is an essential task in cosmology.

One of the most fundamental tests of the Copernican principle comes from observations of our motion with respect to the cosmic microwave background (CMB) rest frame, which induces a kinematic dipole that has already been observed in the CMB \citep{Planck:2013kqc,Planck:2020qil,Ferreira:2020aqa,Saha:2021bay}, the local bulk flow \citep{Colin:2010ds,Feindt:2013pma,Hudson:2004et,Carrick:2015xza}, X-ray clusters \citep{Migkas:2020fza,Migkas:2021zdo}, type Ia supernovae \citep[SNe;][]{Mohayaee:2021jzi,Rahman:2021mti}, high redshift radio sources \citep{Colin:2017juj,Bengaly:2017slg,Siewert:2020krp}, and distant quasars \citep{Secrest:2020has}.\footnote{See also \citep[][]{Aluri:2022hzs}, and references therein, for a recent review of observational tests of the FLRW paradigm.} Many of these observations have intrinsic systematic errors that have to be taken into account \citep{Dalang:2021ruy} in order to avoid theoretical biases. Another route is to perform null tests of the FLRW metric \citep[see][for recent forecasts]{Euclid:2021frk} or to estimate the homogeneity scale \citep[][]{Yadav:2010cc,Kim:2021osl}.

Additionally, it is also possible to test the Copernican principle by assuming an inhomogeneous metric, such as that of the Lemaître-Tolman-Bondi (LTB) model \citep{Garcia-Bellido:2008vdn,February:2009pv,Valkenburg:2012td,Redlich:2014gga}. In fact, current observations can meaningfully test the Copernican principle, leading to constraints on  deviations from  the FLRW metric at almost the cosmic variance level \citep{Camarena:2021mjr}. 

In this paper we explore the precision with which next-generation surveys will probe for violations of the Copernican principle. Specifically, we focus on \Euclid \citep[][]{Laureijs:2011gra}, which is an M-class space mission of the European Space Agency planned to be launched in 2023. The satellite will carry two instruments on board -- the visible imager \citep[VIS;][]{VIS_paper} and the near-infrared spectrophotometric instrument \citep{2012SPIE.8442E..0WP,2016SPIE.9904E..0TM} -- which will carry out a photometric and spectroscopic galaxy survey covering over $15\,000\, \mathrm{deg}^2$ of the sky, with the aim of measuring the growth of the large-scale structure (LSS) up to a redshift of $z\sim 2$~\citep{Euclid:2021icp}.

\Euclid will have two main, complementary cosmological probes, namely galaxy clustering and weak lensing from the photometric survey and galaxy clustering from the spectroscopic survey. While photometric surveys image a larger number of galaxies than spectroscopic ones, they also have larger redshift uncertainties. On the other hand, spectroscopic galaxy surveys have much higher radial precision, but target many fewer objects. \Euclid has very high spectroscopic accuracy; it will be able to make very precise measurements of galaxy clustering that also include the radial dimension, that is to say, it will be able to probe clustering along the line of sight.In this work we create mock baryon acoustic oscillation (BAO) data, in accordance with \Euclid's spectroscopic survey specifications, based on the Fisher matrix approach of \citet[][hereafter EC20]{IST:paper1}

Furthermore, we also stress some of the possible synergies between \Euclid and other contemporary surveys. The latter include the Legacy Survey of Space and Time (LSST) performed at the \textit{Vera C. Rubin} Observatory \citep{Abell:2009aa} and that of the Dark Energy Spectroscopic Instrument \citep[DESI;][]{DESI2016} since they will be complementary to \Euclid in terms of redshift, thus significantly extending the possible redshift range of our analysis.

Finally, forecast constraints on deviations from the Copernican principle were presented in \citet{ReviewDoc}, where a joint analysis between \Euclid \citep{Laureijs:2011gra} and a stage IV SN mission \citep[][assuming SNAP as a concrete example]{DEtaskforce} was performed. Here we update the constraints of this analysis by using more recent \Euclid specifications \citepalias[see][]{IST:paper1} while also considering synergies with other surveys.

This paper is organised as follows: In \cref{sec:model} we
briefly review the dynamics of a spherically inhomogeneous space-time based on the LTB metric but with the addition of a cosmological constant, $\Lambda$ (i.e. the \lltb model) and discuss our particular choices for its arbitrary functions. In \cref{sec:data} we present the data used in our analysis and explain how mock catalogues are produced considering particular fiducial cosmologies, while in \cref{sec:CP} we define and discuss the Copernican prior. Our results are presented and discussed in \cref{sec:results} and \cref{sec:discussion}. We conclude in \cref{sec:conclusions}.

\section{Spherically symmetric inhomogeneous models with a cosmological constant \label{sec:model}} 

A spherically inhomogeneous space-time can be modelled using the $\Lambda$LTB model, which practically is a standard cosmological constant $\Lambda$ and cold dark matter (CDM) model endowed with a spherical inhomogeneity. Here, we aim to test the homogeneity of the Universe, and thus, we neglect anisotropic degrees of freedom by placing the observer at the centre of the spherical inhomogeneity. In this section we briefly review the \lltb model presented  in \citet{Camarena:2021mjr};\footnote{The notation adopted here differs from the notation used in \citet{Camarena:2021mjr}.} a comprehensive review is given in \citet{Marra:2022ixf}.

Hereafter, we use a prime to denote a partial derivative with respect to the radial coordinate, $r$, while we use a dot to denote a partial derivative with respect to the time coordinate, $t$.

\subsection{Dynamics \label{subsec:LTBmetric}}

The LTB metric can be written as
\be
\d s^2 = -c^2 \d t^2 + \frac{R'^2(t,r)}{1-K(r)\,r^2}\d r^2 + R^2(r,t) \, \left( \d\theta^2 + \sin^2 \theta\,\d\phi^2 \right) \,, \label{eq:metric}
\ee 
where the curvature $K(r)$ is an arbitrary function of the radial coordinate. The FLRW metric can be recovered by imposing $K\ = \textrm{constant}$ and $R = a(t)\,r$, with $a(t)$ as the FLRW scale factor. 
From the line element~\Cref{eq:metric}, we can define the transverse and longitudinal scale factors, $\aperp = R(r,t)/r$ and $\apar = R'(r,t)$, respectively. The two scale factors define two different expansion rates given by
\begin{align}
\Hperp(t,r)   \equiv  \frac{\adotperp}{\aperp} \,,
\qquad
\Hpar(t,r)  \equiv  \frac{\adotpar}{\apar} \,.
\end{align}

Solving Einstein's equations with a cosmological constant, we obtain an equation analogue to the first Friedmann equation,
\be
\frac{\Hperp^2}{\Hperpo^2}=\Omega_{\rm m,0} \, \left(\frac{\aperpo}{\aperp}\right)^{3} + \Omega_{k,0} \, \left(\frac{\aperpo}{\aperp}\right)^{2} + \Omega_{\Lambda,0} \,, \label{eq:friedmann}
\ee
where the present-day density parameters are now functions of~$r$,
\begin{align}
\Omega_{\Lambda,0}(r) & = \frac{\Lambda c^2}{3\Hperpo^2 }\,, \label{eq:OmegaLambda} \\
\Omega_{k,0}(r) & = -\frac{K(r)\: c^2}{\Hperpo^2\,\aperpo^2}\,, \label{eq:OmegaK} \\
\Omega_{\rm m,0}(r) & = \frac{2\,m(r)}{\Hperpo^2 \,\aperpo^3 \,r^3}\,, \label{eq:OmegaM} 
\end{align}
which satisfy $\Omega_{\rm m,0}(r)+ \Omega_{k,0}(r)+\Omega_{\Lambda,0}(r)=1$. It should be noted that we have defined $\Hperpo \equiv \Hperp(t_0,r)$ and $\aperpo \equiv \aperp(t_0,r)$. 

From~\Cref{eq:OmegaM} we can see that Einstein's equations introduce another arbitrary function: the Euclidean mass $m(r)$. This function arises as a constant of integration and is defined via
\be
m(r)  = \int_0^{r} \d r' \: 4 \pi G \, \rho_{\rm m}(t,r') \, a_{\parallel}\, a_{\perp}^2 r'^2\,, \label{eq:rhom}
\ee
where $\rho_{\rm m}(t,r)$ is the local matter density.

Another arbitrary function of the \lltb model is the Big Bang function, $t_{\rm BB}(r)$. This can be interpreted as the time corresponding to the Big Bang singularity surface, and it emerges from integration of~\Cref{eq:friedmann}:
\be 
t - t_{\rm BB}(r) = \frac{1}{\Hperpo(r)}\int_{0}^{\chi} \frac{\d  x}{\sqrt{\Omega_{\rm m,0} (r)x^{-1} + \Omega_{k,0} (r) +  \Omega_{\Lambda,0}(r) x^{2}}} \,, \label{eq:tbb}
\ee
where $\chi = \aperp (t,r)/ \aperpo$. We note that the last integral defines the age of the Universe, $t_0$, when integrated from zero to one.  

Finally, from the line element \Cref{eq:metric} it follows that the geodesic equations are
\begin{align} \label{eq:geodesics}
\frac{\d t}{\d z}  = - \frac{1}{(1+z)\,\Hpar(t,r)} \,,
\qquad
\frac{\d r}{\d z}  = - \frac{c\sqrt{1 - K(r)r^2} }{(1+z)\,\apar(t,r)\,\Hpar(t,r)} \,.
\end{align}

\subsection{Free functions \label{subsec:profile}}

As shown above, the \lltb model has three arbitrary functions: $K(r)$, $m(r)$, and $t_{\rm BB}(r)$. One of these functions is just a gauge freedom that can be fixed by re-scaling the radial coordinate. Here, we fix the radial coordinate $r$ such that the mass function satisfies $m(r) \propto r^3$; we will later present the missing normalisation  of the Euclidean mass. On the other hand, a Big Bang function different from zero introduces decaying modes into the matter density \citep{Silk:1977vv}. The presence of these modes leads to a disagreement with the standard scenario of inflation~\citep{Zibin:2008vj}. Thus, to ensure the absence of decaying modes on the matter density we set $t_{\rm BB}(r)=0$, which also implies that the Big Bang singularity happens everywhere simultaneously.

Hence, we end up with just one arbitrary function, the curvature profile $K(r)$. Here, we adopt the monotonic compensated profile given by
\be 
K(r)= K_{\rm B} + (K_{\rm C} - K_{\rm B}) \, P_3 (r/ r_{\rm B} ) \,, \label{eq:kr}
\ee
where $r_{\rm B}$ is the comoving radius of the spherical inhomogeneity, $K_{\rm B}$ is the background curvature, $K_{\rm C}$ is the central curvature, and the function
$P_3$ follows,
\begin{align}
P_{3}(x)= \left\{\begin{array}{ll}
1 - \exp \big[-(1-x)^3/x] & \mbox{ for }  0  \le x < 1\\
0 & \mbox{ for } 1 \leq x 
\end{array}\right. \,, 
\end{align}
with $x=r/ r_{\rm B}$. Thus, \Cref{eq:kr} ensures a smooth transition between the LTB and FLRW metrics; the \lltb model asymptotes to the \lcdm model at scales $r \geq r_{\rm B}$. 

We note that while we  have fixed $m(r)$ and $t_{\rm BB}(r)$ using physical arguments, our choice of $K(r)$ remains arbitrary and could have a significant impact on our analysis. In \Cref{app:kr} we discuss our choice and compare it with another \lltb model. We also present an extra analysis performed using an extension of \Cref{eq:kr}. This further analysis shows that a more generalised curvature profile could weaken the constraints by up to a factor of $\sim 2.$

Once the three arbitrary functions are fixed, we can compute $\rho_{\rm m}(r,t)$ in order to determine the matter density contrast using
\be
\delta (r,t) :=  \frac{\rho_{\rm m} (r,t)}{ \rho_{\rm m} (r_{\rm B},t)} - 1 \label{eq:deltam} \,,
\ee
and we can also compute the mass (integrated) density contrast via
\begin{align}
 \Delta (r, t_0) &= \frac{4\pi \int_0^r \d \bar r \, \delta(\bar r,t_0)\,  \aperp^2(\bar r,t_0) \apar(\bar r,t_0) r^2 }{4\pi\, \aperp^3(r,t_0)\,r^3/3} \label{eq:Deltar} \\
& =  \frac{m(r)}{4\pi G\, R^3(r,t_0) /3 \, \rho_{\rm m}^{\rm{out}}(t_0)} -1 
 = \frac{\Omega_{\rm m,0}(r)}{\Omega_{\rm m,0}^{\rm{out}}} \left[ \frac{H_{\perp 0}(r)}{H_{\perp 0}^{\rm{out}}} \right]^2 -1 \,. \nonumber
\end{align}
Here we use the superscript `out' to denote the FLRW background quantities outside the inhomogeneity, for example~$\rho_{\rm m}(r_{\rm B},t_0)=\rho_{\rm m}^{\rm{out}}(t_0)$. We additionally make use of the FLRW comoving coordinate at the present time, which is defined as
\be
r^{\rm out} := r\:\aperpo(r)/a^{\rm out}(t_0) \,. \label{eq:rout}
\ee

\begin{figure}[!t]
\centering
\includegraphics[clip, width = \columnwidth]{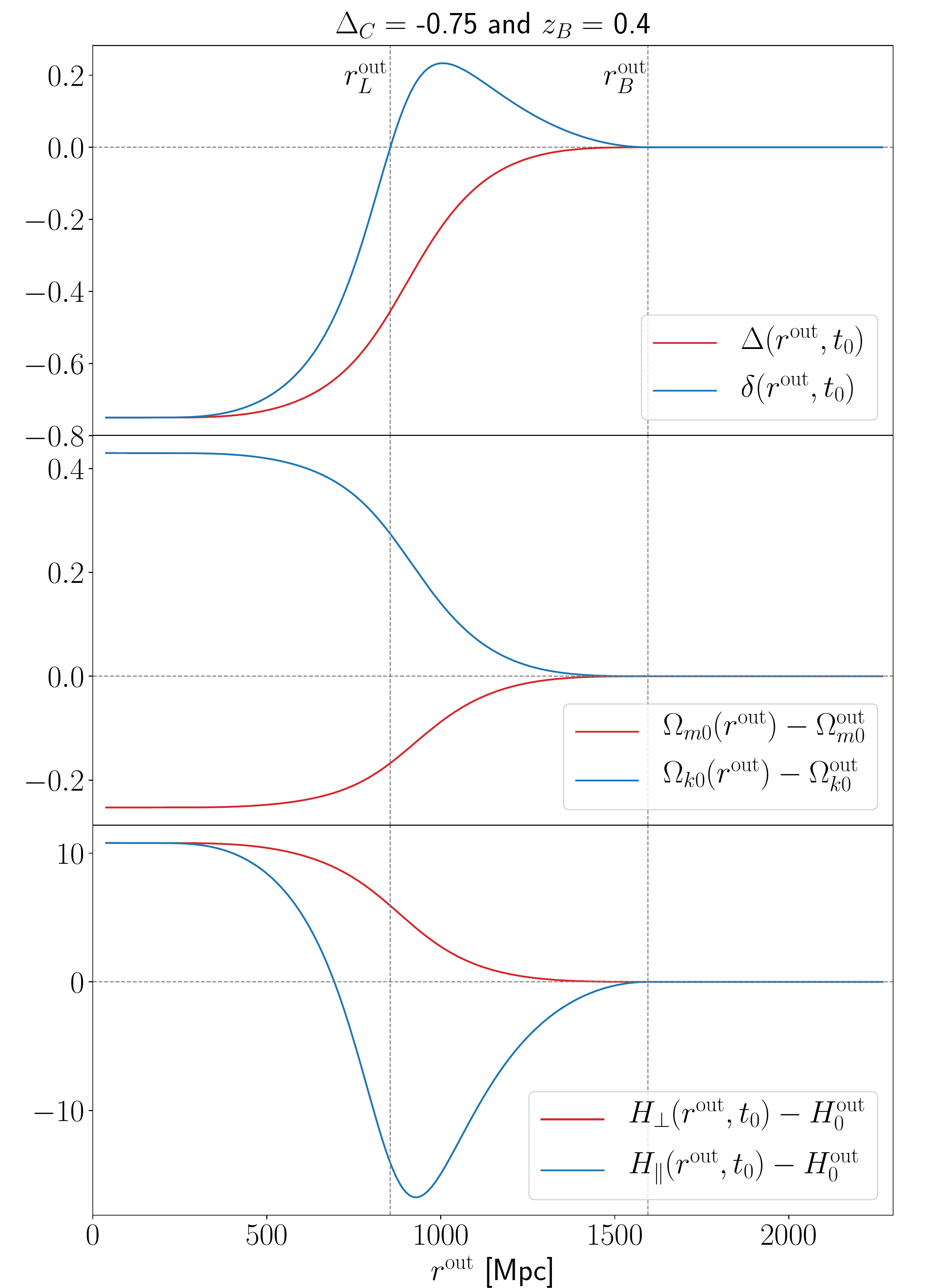}
\caption{Several \lltb quantities as a function of the FLRW comoving coordinate $r^{\rm{out}}$. Top: Density contrast of matter, $\delta (r) = \rho_{\rm m}(r,t_0)/ \rho_{\rm m}(r_{\rm B},t_0) - 1$, and integrated mass, $\Delta (r)$, as functions of the FLRW comoving coordinate. At radius $r_{\rm L}^{\rm out}$ the central structure ends and the compensating shell begins, while $r_{\rm B}^{\rm out}$ is the boundary of the spherical inhomogeneity. Middle: Deviations from the background density of matter, $\smash{\Omega^{\mathrm{out}}_{\rm m,0}}$, and curvature, $\smash{\Omega^{\mathrm{out}}_{k,0}}$, as a function of the FLRW comoving coordinate due to the radial dependence on $\Omega_{\rm m,0}(r)$ and $\Omega_{k,0}(r)$. We note that the FLRW background quantities are recovered for $r \ge r^{\mathrm{out}}_{\rm B}$. Bottom: Transverse and longitudinal fluctuations in the expansion rates as a function of the FLRW comoving coordinate.\label{fig:density}}
\end{figure}

The top panel in \Cref{fig:density} shows the matter and integrated mass density contrast as a function of $r^{\rm out}$  at $t=t_0$ for a deep and gigaparsec-scale void. Also displayed in the figure are $\Omega_{\rm m,0}(r^{\mathrm{out}})-\Omega^{\mathrm{out}}_{\rm m,0}$ and $\Omega_{k,0}(r^{\mathrm{out}})-\Omega^{\mathrm{out}}_{k,0}$, giving the deviations of the matter and curvature densities with respect to their \lcdm counterparts (middle panel), along with the deviations of the transverse and longitudinal expansion rates with respect to $H_0^{\rm out}$ (bottom panel).

It is important to highlight that the assumption of the profile \Cref{eq:kr} implicitly introduces a compensating scale, here denoted by $r^{\rm out}_{\rm L} [\delta(r^{\rm out}_{\rm L},t) = 0 ]$, at which the central overdense or underdense region makes a transition to the surrounding mass-compensating underdense or overdense region. The $r^{\rm out}_{\rm L}$ and the $r^{\rm out}_{\rm B}$ scales are represented in \Cref{fig:density} by the dotted vertical lines. Furthermore, one can note that at the centre of the inhomogeneity $\Delta (0,t) = \delta (0,t)$, while at the boundary shell we have $\Delta (r_{\rm B},t) = \delta (r_{\rm B},t) = 0$. Since the LTB and FLRW metrics perfectly match at the boundary shell, then we have that $r^{\rm out}_{\rm B} = r_{\rm B}$. Finally, from~\Cref{eq:OmegaM} is possible to determine the missing normalisation  of the Euclidean mass; specifically we find $m(r) = \Omega_{\rm m,0}^{\rm{out}}\,(H_{0}^{\rm{out}})^2\,r^3/2$, while \Cref{eq:OmegaK} leads to $K_{\rm B} = - \Omega_{k,0}^{\rm{out}}\,(H_{0}^{\rm{out}})^2$.

\subsection{Parameter space \label{subsec:free}}

We used the \texttt{monteLLTB} code to solve the dynamical equations and then to sample the parameter space of the \lltb model.\footnote{\url{https://github.com/davidcato/monteLLTB}} The \texttt{monteLLTB} code combines \texttt{montepython} \citep{Audren:2012wb,Brinckmann:2018cvx} for the Markov chain Monte Carlo parameter space exploration and likelihoods, \texttt{class} \citep{Blas:2011rf} for the CMB computation  and \texttt{voiddistances2020} \citep{Valkenburg:2011tm} for the $\Lambda$LTB metric functions via a wrapper that translates  the \texttt{montepython} trial vector into an effective FLRW vector that is suitable for \texttt{class} (see \citealt{Camarena:2021mjr} for details).

Since the LTB metric asymptotes to FLRW at~$r \ge r_{\rm B}$, the background expansion of our model is specified by the standard six $\Lambda$CDM parameters: the normalised Hubble constant $h := H_0/100$, the baryon density $\Omega_\mathrm{b,0}$, the cold dark matter density $\Omega_\mathrm{cdm,0}$, the optical depth $\tau$, the amplitude of the power spectrum $A_{\rm s}$ and its tilt $n_{\rm s}$.

On the other hand, the spherically inhomogeneous region is fixed by the boundary redshift, $z_{\rm B}$, and the mass density contrast at the centre of the inhomogeneity, $\Delta_{\rm C} \equiv \Delta(0,t) $. The parameters $z_{\rm B}$ and $\Delta_{\rm C}$ are related to the parameters $r_{\rm B}$ and $K_{\rm C}$, which explicitly appear in the definition of the curvature profile. As discussed in \Cref{sec:CP}, we conveniently present our results using the compensating scale $r_{\rm L}^{\rm out}$ and the mass contrast at such a scale, $\Delta_{\rm L} \equiv \Delta(r_{\rm L},t_0)$.

We note that for $\Delta_{\rm C}\,\approx\,0$, the parameter space is highly degenerate so that the model could feature arbitrarily large values of $z_{\rm B}$ and be still allowed by the data. In order to overcome this issue, we adopt, as in  \cite{Camarena:2021mjr}, the flat prior $z_{\rm B} \in [0, 0.5]$. This prior allows us to map non-Copernican structures of up to $r_{\rm L}^{\rm out} \sim 1 $ Gpc.

\section{Data \label{sec:data}}

Here we present both the forecast and current data used to constrain the \lltb model. Although our goal is to forecast constraints on the Copernican principle given the forthcoming surveys, the inclusion of current data is needed to tightly constrain the \lltb parameter space at small scales.

Forecasted datasets are generated considering four fiducial models, which are based on the \lcdm and \lltb models (see \Cref{tab:fiducials}). By considering the \lcdm forecast data we aim to determine the precision with which next-generation surveys will be able to probe for deviations from the FLRW metric. Meanwhile, by considering the inhomogeneous \lltb fiducial model, we aim to investigate the ability of future surveys to detect a violation of the Copernican principle. In order to assume a consistent fiducial model, current data have been re-scaled to agree with the aforementioned fiducial models. Such a re-scaling is performed following the procedure described in \Cref{app:mock_LLTB}. Furthermore, we also assume that there are no tensions among the different datasets; this includes tensions between early and late determinations.
The theoretical predictions for the observables implemented here follow the equations discussed in Sect.~3 of \cite{Camarena:2021mjr}. We note that, in order to fully calibrate the SN distances, we also assume a fiducial value for the absolute magnitude of SN data, that is, $M_0 = -19.3$.

\subsection{Forecast data \label{sub:forecast}}

In order to forecast how \Euclid and other forthcoming surveys will constrain deviations from the Copernican principle, we created mock SN and BAO data; the former provide information on the luminosity distance, while the latter concern the Hubble parameter and the angular diameter distance. In particular, the recipes described here were used for the \lcdm catalogues, while the \lltb catalogues are obtained by suitably re-scaling the former (see \Cref{app:mock_LLTB}).

\begin{table}
\begin{center}
\setlength{\tabcolsep}{4pt}
\renewcommand{\arraystretch}{1.35}
\caption{Parameter values for the fiducial models that are used for the mocks. The values used for the \lcdm follow the fiducial of \citetalias{IST:paper1}; in particular, spatial flatness is assumed. The Hubble constant $H_0$ is shown in units of km\:s$^{-1}$\:Mpc$^{-1}$, while the absolute magnitude of the SN $M_0$ is shown in units of mag. \label{tab:fiducials}}
\begin{tabular}{lcccccccc}
\hline
\hline
model & $M_0$ & $\Omega_{\rm m,0}$ & $\Omega_{\rm b,0}h^2$ & $H_0$ & $\Delta_{\rm C}$ & $z_{\rm B}$\\ \\
 \hline
\lcdm &$-19.3$ & $0.32$ & $0.02225$ & $67$ & - & -\\
\lltb $1$ &$-19.3$ & $0.32$ & $0.02225$ & $67$ & $-0.5$ & $0.05$ \\
\lltb $2$ &$-19.3$ & $0.32$ & $0.02225$ & $67$ & $-0.1$ & $0.4$ \\
\lltb $3$ &$-19.3$ & $0.32$ & $0.02225$ & $67$ & $-0.1$ & $0.8$ \\
\hline
\hline
\end{tabular}
\end{center}
\end{table}
The four fiducial cosmologies based on the \lcdm and the \lltb model that we consider here are shown in \Cref{tab:fiducials}; the former was also used in \citetalias{IST:paper1}.
To make the mocks for the \lcdm model, we calculated the redshift evolution of the Hubble parameter, along with the luminosity and angular diameter distances, using the recipe described in the next section, which is based on the specifications of \Euclid and other LSS surveys. On the other hand, as mentioned before, the \lltb mock catalogues are obtained following the process described in \Cref{app:mock_LLTB}. Since computing correlation matrices for models far from the \lcdm model is currently not possible \citep{Harnois-Deraps:2019rsd,DES:2020ypx,Ferreira:2021dmb}, we first compute the correlation matrix assuming the \lcdm model. Then, we apply the method described in \Cref{app:mock_LLTB} to obtain the corresponding \lltb matrices.

Since future surveys like \Euclid are expected to provide observations with high precision, it is important to be convinced that our analysis methods will be robust. Hence, in order to understand and take possible observational systematic uncertainties that can affect the measurements into account, several analyses, such as that of \citet{Paykari2020}, have been performed. In the latter, the observational systematic effects of the \Euclid VIS instrument were studied, taking the modelling of the point spread function and the charge transfer inefficiency into account. Since these systematic effects are expected to be better understood by the time the data arrive, in this analysis we assume that they will be under control in the final data products. In any case, in what follows we in fact include several astrophysical systematic effects, such as the galaxy bias, as discussed in what follows.

\subsubsection{SN surveys}

In our analysis we focus on two forthcoming SN surveys, the first of which is based on the proposed \Euclid DESIRE survey \citep{Laureijs:2011gra,Astier:2014swa}, while the second one is based on the specifications of the LSST. In particular, we assume that the \Euclid DESIRE survey will observe $1700$ SNe in the redshift range $z\in[0.7,1.6]$, while the one from the LSST survey will observe $8800$ SNe in the redshift range $z\in[0.1,1.0]$, thus resulting in a total of $10\,500$ points.

In either case we consider the redshift distributions of the SN events as described in \citet{Astier:2014swa}, assuming the points are not correlated with each other. Even though the \Euclid SN survey is not currently guaranteed to take place, we decided to include it in order to extend the redshift range of LSST at high~$z$. For the SN mocks we include an observational error of the form
$\sigma_{\textrm{tot},i}^2 = \delta \mu^2_i + \sigma^2_{\textrm{flux}} + \sigma^2_{\textrm{scat}} + \sigma^2_{\textrm{intr}}$,
where the terms corresponding to the intrinsic contributions, the scatter and the flux are the same for all events: $\sigma_{\textrm{intr}} = 0.12$, $\sigma_{\textrm{scat}} = 0.025$, and $\sigma_{\textrm{flux}} = 0.01,$ respectively. Finally, we also include an error on the distance modulus $\mu=m-M_0$ that scales linearly with $z$ as $\delta\mu = e_{\rm M}~z$,
where  $e_{\rm M}$ follows a Gaussian distribution with zero mean and standard deviation $\sigma(e_{\rm M})=0.01$ \citep[see][]{Gong:2009yk,Astier:2014swa}, which includes the possible redshift evolution of SNe not taken into account by the distance estimator (see \citealt{Astier:2014swa}). However, while a value of $e_{\rm M}=0.01$ is required to take a possible systematic evolution into account, this would be added quadratically to an effective term of $e_{\rm M}=0.055$ arising from SN lensing. The latter has been theoretically calculated by several authors to be of the order of $\sigma_{\rm lens} \simeq 0.055\,z$, for example $\sigma_{\rm lens}= 0.052\,z$ \citep[][]{Amendola:2013twa,Quartin:2013moa}, and $\sigma_{\rm lens}= 0.056 \,z$ \citep[][]{Ben-Dayan:2013nkf}, while observationally it was determined, via the Supernova Legacy Survey to be $\sigma_{\rm lens}= (0.055 \pm 0.04)\,z$ \citep[][]{Jonsson:2010wx} and $\sigma_{\rm lens}= (0.054 \pm 0.024)\,z$ \citep[][]{SNLS:2010rmd}.

\subsubsection{Local prior on the Hubble constant}

We also forecast a $1\%$ measurement of the Hubble constant, which is the grand goal of the SH0ES collaboration,
\begin{align} 
H_0 & =
\begin{cases}
67.00 \pm 0.67\,  \kmsMpc 
&\textrm{for}\  \Lambda\textrm{CDM} \\[1mm]
67.62 \pm 0.68\,  \kmsMpc &\textrm{for}\  
\Lambda\textrm{LTB}~1 \\[1mm]
68.22 \pm 0.68\,  \kmsMpc &\textrm{for}\  
\Lambda\textrm{LTB}~2 \\[1mm]
68.45 \pm 0.68\,  \kmsMpc &\textrm{for}\  
\Lambda\textrm{LTB}~3
\end{cases},
\label{eq:H0}
\end{align}
where the central value is given by the fiducial $H_0$ value for the \lcdm fiducial model, meanwhile for the \lltb models this central value is the expected value given the methodology described in \Cref{app:HL}.
Here, as mentioned earlier, we consider a scenario in which there is no tension between early and late determinations of the Hubble constant. By assuming a single consistent fiducial model, we focus on the constraining potential of future surveys to test the Copernican principle, leaving the issue of the Hubble tension to other studies. This is in part justified since \citet{Camarena:2021mjr,Camarena:2022iae} shown that a large inhomogeneity cannot explain away the Hubble tension. 
Finally, we impose the Gaussian prior of \Cref{eq:H0} on $H_0^{\rm L}$; the corresponding Hubble constant value for an inhomogeneous model (see \Cref{app:HL} for a detailed discussion).

\subsubsection{Large-scale structure surveys}

Here, we now briefly describe our procedure for creating mock BAO data based on the specifications of \Euclid via a Fisher matrix approach, following the methodology of \citetalias{IST:paper1} for the spectroscopic survey, on which we focus since we are interested in obtaining precise measurements of the angular diameter distance $D_\mathrm{A}(z)$ and the Hubble parameter $H(z)$.
We do not consider weak lensing by \Euclid, nor other perturbation level observables such as redshift space distortions, because there is not yet a fully developed linear perturbation theory on inhomogeneous backgrounds such as the LTB.
\footnote{See, however, \citet{Moss:2010jx,Ishak:2013vha} for a comparison with observations.}
A discussion and the numerical simulation of the LSS on an LTB background is provided in \citet{Marra:2022ixf} and references therein.

As was extensively discussed in \citetalias{IST:paper1}, the main targets of the Euclid survey will be emission line galaxies (ELGs), which are bright emitters in specific lines, such as $H_\alpha$ and [O III], that can be seen in the redshift range $z \in [0.9,1.8]$, and can be used to measure the galaxy power spectrum. In particular, \Euclid will determine approximately $30$ million spectroscopic redshifts with an uncertainty of $\sigma_z = 0.001(1 + z)$ \citep{Pozzetti:2016cch}, which will provide the galaxy power spectrum with information on the distortions due to the redshift uncertainty, the residual shot noise, the Alcock-Paczynski effect, the redshift space distortions and the galaxy bias. Furthermore, non-linear effects, such as a non-linear smearing of the BAO feature or a non-linear scale-dependent galaxy bias that distorts the shape of the power spectrum, have also been taken into account (see \citealt{Wang:2012bx} and \citealt{delaTorre:2012dg}, respectively).

In this work we again make use of the same binning scheme as in \citet{Martinelli:2020hud, Euclid:2021cfn}, which differs from that of \citetalias{IST:paper1}. In particular, instead of four equally spaced redshift bins, we now consider nine bins of width $\Delta z = 0.1$. After re-binning the data provided in \citetalias{IST:paper1}, we obtain the following specifications for the galaxy number density $n(z)$, given in units of $\mathrm{Mpc}^{-3}$, and that of the galaxy bias $b(z)$:
\begin{align}
n(z)& \!\!=\!\! \{2.04, 2.08, 1.78, 1.58, 1.39, 1.15, 0.97, 0.7, 0.6\}\!\times\! 10^{-4}\!\!, \\
b(z)&\!=\! \{1.42, 1.5, 1.57, 1.64, 1.71, 1.78, 1.84, 1.90, 1.96\} \,.
\end{align}
In \citet{Martinelli:2020hud} we tested our choice for the binning scheme against that of \citetalias{IST:paper1}, and we found the results were in agreement.

In the case of the \lcdm mocks, the Fisher matrix for the cosmological parameters, along with the associated covariance matrix, can be derived by following the methodology described in \citetalias{IST:paper1}. The cosmological parameters we consider for the \lcdm mocks include the background quantities $\{\omega_{\rm m}=\Omega_{\rm m,0}h^2$, $h$, $\omega_{\rm b}=\Omega_{\rm b,0}h^2$, $n_{\rm s}\}$, two non-linear parameters $\{\sigma_{\rm p},\,\sigma_{\rm v}\}$ (see \citetalias{IST:paper1}) and the five redshift-dependent parameters $\{\ln D_{\rm A},\,\ln H,\,\ln f\sigma_8,\,\ln b\sigma_8,\,P_{\rm s}\}$, which are estimated in every redshift bin. Here we have defined $f\sigma_8\equiv f(z)\sigma_8(z)$ as the linear growth rate multiplied by $\sigma_8$, which corresponds to the RMS fluctuations in the matter mass density in a comoving sphere of $8\,h^{-1}\,\mathrm{Mpc}$, while $b\sigma_8\equiv b(z)\sigma_8(z)$ and $P_{\rm s}$ are the galaxy bias and the shot noise, respectively (see \citetalias{IST:paper1}). 
From this we can then estimate the expected uncertainty of the measurements of the Euclid survey for both the angular diameter distance $D_{\rm A}(z)$ and the Hubble parameter $H(z)$, in every redshift bin, while all other parameters are marginalised  over. Furthermore, we apply the approach presented in \Cref{app:mock_LLTB} to obtain the corresponding \lltb mock data.

Since the spectroscopic survey of \Euclid will only cover the redshift range $z \in [0.9,1.8]$, this limits the range where SN and BAO data will be obtained. Hence, in order to cover smaller redshifts we complement our analysis by using fiducial data products from the DESI survey as well. DESI has already initiated survey operations in 2021 and will eventually obtain spectra for tens of millions of galaxies and quasars up to $z\sim 4$, thus making redshift-space distortion and BAO analyses possible.
To create DESI mocks, assuming the \lcdm model, we follow the methodology for both the angular diameter distance $D_{\rm A}(z)$ and the Hubble parameter $H(z)$, as described in \citet{DESI2016}. These Fisher matrix forecasts were also derived using the full anisotropic galaxy power spectrum (i.e. measurements of the matter power spectrum as a function of the angle with respect to the line of sight), as described in \citet{2014JCAP...05..023F}. This approach is similar to that of the \Euclid forecasts and it also includes all information from the two-point correlation function.
In particular, the baseline DESI survey will cover approximately $14\,000\,\mathrm{deg}^2$ and will target ELGs, luminous red galaxies, bright galaxies, and quasars, all in the redshift range $z\in [0.05,3.55]$, although the precision of the measurements will depend on the target population. Regarding the specific populations, the bright galaxies will be in the range $z\in [0.05,0.45]$ in five equally spaced redshift bins, while the ELGs and the luminous red galaxies will be in the range $z\in [0.65,1.85]$ in $13$ equally spaced bins. Finally, the Ly-$\alpha$ forest quasars will be in the range $z\in [1.96,3.55]$ in $11$ equally spaced bins and we assume that the points are uncorrelated with each other.

In the case when we used the combination of \Euclid and DESI data together, in order to avoid overlap between the two surveys at late times, we only considered the DESI points that do not overlap with those of \Euclid, because an overlap will lead to undesired correlations between the surveys. Moreover, since the DESIRE + LSST SN points will only reach at most $z=1.6$, we included the DESI data up to $z=0.9$, thus omitting the Ly-$\alpha$ forest observations. However, when used separately we considered the full redshift range of the datasets.

\subsection{Current data \label{sub:current}}

As shown in \cite{Camarena:2021mjr}, CMB and SN data are necessary in order to obtain tight constraints on the \lltb model. For our particular case, this means that the presence of real data (i.e. \Planck 2018 and Pantheon SNe) is needed even though our analysis aims to forecast the contribution of forthcoming surveys. The inclusion of CMB data is crucial to constrain the background parameters, while the usage of low-$z$ SNe allows us to break the degeneracy of the \lltb parameters model at small scales. As discussed at the beginning of the present Sect., we rescale current data according to the predictions of the fiducial models shown in \Cref{tab:fiducials} and following the procedure described in \Cref{app:mock_LLTB}.

\subsubsection{Cosmic microwave background}

When the \lcdm forecast data are considered, we perform our analysis including the latest \Planck CMB data\footnote{\href{http://www.esa.int/Planck}{http://www.esa.int/Planck}} \citep{Planck:2018vyg}. We use the high-$\ell$ TT+TE+EE, low-$\ell$ TT, and low-$\ell$ EE likelihoods. Particularly, we use the compressed version of high-$\ell$ data, that is, the likelihood normalised over all nuisance parameters except $A_\mathrm{planck}$. We note that typical constraints obtained for $\Lambda$CDM using these likelihoods include the fiducial values adopted for the forecast data (\Cref{tab:fiducials}) within $68\%$ uncertainties, allowing the combination of CMB and forecast data without the necessity of applying the re-scaling technique. 

On the other hand, the \lltb cosmologies presented on \Cref{tab:fiducials} could significantly change the CMB power spectra and lead to disagreements of these with the constraints of the aforementioned likelihoods. Thus, one should change the \Planck CMB data according to the \lltb fiducial cosmologies. This is not a trivial task given the complex structure of the CMB likelihoods and our limited understanding of perturbations on the inhomogeneous models. Thus, for our analyses of the \lltb mock data we use the CMB distance priors on the shift parameter $R$, the acoustic scale $l_A$, the amount of baryons $\Omega_{\rm b} h^2$, and the tilt of the power spectrum $n_{\rm s}$. We build the mock CMB priors considering the current measurements given by \cite{Chen:2018dbv}.

\subsubsection{SN surveys}

The lack of SN data at very low redshifts $z \sim 0.01$ -- the lowest LSST point lies at $ z = 0.1$ -- increases the degeneracy between $\Delta_{\rm C}$ and $z_{\rm B}$, loosening the constraints on the $\Lambda$LTB model. To overcome this issue, we include the Pantheon SN compilation \citep{Scolnic:2017caz}.

\subsubsection{Large-scale structure surveys}

We also include BAO data from 6dFGS \citep{Beutler:2011hx}, SDSS-MGS \citep{Ross:2014qpa} and BOSS-DR12 \citep{Alam:2016hwk} surveys. The isotropic measurements from 6dFGS and SDSS-MGS allow us to access redshifts 0.1 and 0.15, respectively, while BOSS  provides anisotropic measurements at redshifts 0.38, 0.51, and 0.61. We note that these current data overlap with our forecast DESI catalogues, but we assume no correlations between these datasets. Hereafter we collectively refer to this set of data as BAOs. We note that our analysis does not include the latest eBOSS data \citep{Ross:2020lqz,Raichoor:2020vio,Lyke:2020tag,duMasdesBourboux:2020pck} chiefly because the eBOSS dataset spans over all the redshift range of our forecast \Euclid data.

\subsubsection{$y$-Compton distortion and the kSZ effect}

Finally, when forecast data from \lcdm were analysed, we introduced priors on the $y$-Compton distortion and the kinetic Sunyaev-Zeldovich (kSZ) effect. For the $y$-Compton distortion, we adopted the upper limit prior at $95.4\%$ uncertainty provided by COBE-FIRAS $y < 1.5 \times 10^{-5}$ \citep{Fixsen:1996nj}. Meanwhile, for the kSZ effect we adopted the $\sim 47\%$ constraint from  SPT-SZ and SPTpol surveys \citep{Reichardt:2020jrr}. Considering the \lcdm fiducial, we implemented the Gaussian prior on the kSZ amplitude as $D_{3000} = \SI{3.49 \pm 1.63} {\micro\kelvin}$.

Priors on the $y$-Compton distortion and the kSZ effect were not used for our analysis of the \lltb forecast data since they do not improve upon constraints given by the combinations of the other datasets. 

\section{Copernican prior \label{sec:CP}}

\begin{figure*}[!th]
\centering
\includegraphics[width =\textwidth]{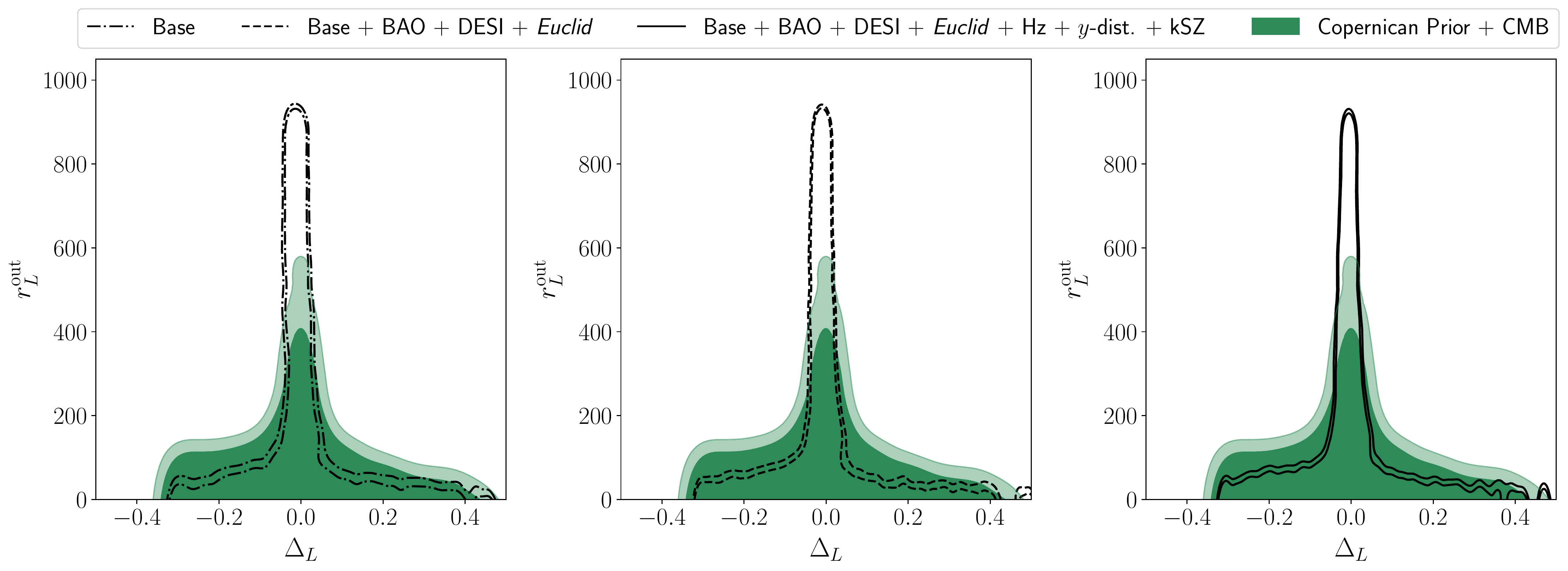}
\caption{$95\%$ and $99\%$ confidence level constraints on the integrated mass contrast, $\Delta_{\rm L}$, and the comoving size, $r^\textrm{out}_{\rm L}$, for three different data combinations as compared to the constraints  from the Copernican prior convolved with the CMB likelihoods. \label{fig:CP}}
\end{figure*}

In the absence of the Copernican principle, the LSS of the Universe may feature arbitrary radial inhomogeneities. In an FLRW model, instead, those structures are constrained by the Copernican principle. Such constraints can be obtained through linear perturbation theory. Assuming that the density contrast $\Delta(r)$ is a Gaussian field, we can compute its RMS by
\begin{equation} \label{eq:sigmar}
\sigma^2(r) = \int_0^{\infty} \frac{\d k}{k} \: \left[ \frac{k^3 P_{\rm m,0}(k)}{2\pi} \frac{3j_1(rk)}{rk} \right]^2 \,,
\end{equation}
where $P_{\rm m,0}(k)$ is the standard power spectrum today and $j_1$ is the spherical Bessel function of the first kind. The aforementioned quantities can be used to define a prior that establishes the probability of finding an inhomogeneous deviation from the FLRW at a given scale.
Such a prior is the so-called Copernican prior and can be used to constrain $\Delta_{\rm C}$ and $z_{\rm B}$ through \citep{Camarena:2021mjr}
\begin{equation} \label{eq:CP}
\mathcal{P}(\Delta_{\rm C},z_{\rm B}) \propto \exp{\left[ -\frac{1}{2} \frac{\Delta^2(r_{\rm L},t_0)}{\sigma^2(r_{\rm L}^{\rm out})} \right]} \,,
\end{equation}
where $\Delta(r,t_0)$ is given by~\Cref{eq:Deltar}, $r_{\rm L}(\Delta_{\rm C},z_{\rm B})$ is the radius of the central under/overdensity, and $r_{\rm L}^\mathrm{out}(\Delta_{\rm C},z_{\rm B})$ is the latter radius in the FLRW comoving coordinates of~\Cref{eq:rout}. We note that $r_{\rm L}^\mathrm{out}(\Delta_{\rm C},z_{\rm B})$ is the scale of interest since it defines the size of the central under/overdensity.
Additionally, by definition the Copernican prior vanishes at the matching shell, $r_{\rm B}^\mathrm{out}$, since the matter and mass fluctuations disappear. We present our results using the radius $r_{\rm L}^\mathrm{out}$ and the mass contrast $
\Delta_{\rm L} \equiv \Delta(r_{\rm L},t_0)$.

Despite the fact that \Cref{eq:CP} can constrain the deviations from the FLRW model by constraining $\Delta_{\rm C}$ and $z_{\rm B}$, this prior does not constrain the cosmological parameters needed to assess the information contained in perturbations, for instance $P_{\rm m,0}$. On the other hand, CMB observations should describe the early Universe at any point and, in particular, also at our observing position if the Copernican principle is valid. That is, under the assumption of the Copernican prior, CMB information such as the power spectrum should constrain $\Delta_{\rm C}$ and $z_{\rm B}$ (and the background cosmological parameters). 

We then compared the cosmological constraints on $\Lambda$LTB with the ones from the Copernican prior convolved with the CMB likelihood to obtain $P$, the probability distribution of $\Delta_{\rm C}$ and $z_{\rm B}$, given the initial conditions obtained from the CMB and their uncertainty, which, under the Copernican principle, describe matter perturbations around us:
\begin{equation} \label{PCMB}
  P( \Delta_{\rm C} , z_{\rm B})  =  \int \d p_i\, \mathcal{P}(\Delta_{\rm C},z_{\rm B}) \, \mathcal{L}_{\rm CMB}(p_i,\Delta_{\rm C},z_{\rm B}) \,,
\end{equation}
where $p_i$ denotes the standard $\Lambda$CDM parameters and $\mathcal{L}_{\rm CMB}$ is the CMB likelihood of \Cref{sub:current}.

\section{Results \label{sec:results}}
As mentioned in \Cref{subsec:free}, we explore the parameter space using the \texttt{monteLLTB} code: a cosmological solver and sampler for the $\Lambda$LTB model. Most of the plots shown in this Sect. have been produced using \texttt{getdist} \citep{Lewis:2019xzd}.

Specifically, we constrained the $\Lambda$LTB model using several combinations of current and forecast data. We defined as a baseline analysis (hereafter `Base') the combination of CMB, Pantheon SN, LSST, and $H_0$ data; however, we neglected possible correlations between LSST and Pantheon. We also defined the baseline analysis relative to current data (hereafter `Base C') as the combination of CMB, Pantheon, and  $M_{\rm B}$ data, with the last being the B-band absolute magnitude of SNe as inferred by the Cepheid distances \citep[see][]{Camarena:2022iae}. We neglected any possible correlation between the future DESI and \Euclid dataset with the current BAOs. When DESI and \Euclid data are combined, we replaced DESI measurements between $z \in [0.95,1.75]$ with the \Euclid data points.

\begin{table}[!th]
\begin{center}
\setlength{\tabcolsep}{8pt}
\renewcommand{\arraystretch}{1.35}
\caption{Ratios of the areas of the $95\%$ contours from observations and the Copernican principle (see \Cref{fig:CP}). We also include (last row) the case with background curvature, $K_{\rm B} \neq 0$ in~\Cref{eq:kr}.\label{tab:Area}}
\small{\begin{tabular}{lcc}
\hline
\hline
\multirow{2}{*}{Observables} & \multicolumn{2}{c}{$A_{\rm obs}$/$A_{\rm CP}$} \\
    & $0 \!\leq\! r^{\textrm{out}}_{\rm L}$  & $190\,\mathrm{Mpc} \!\leq\! r^{\textrm{out}}_{\rm L} $ \\ \hline
\hline
\multicolumn{3}{c}{Flat background FLRW metric} \\ \hline
Base (CMB+Pantheon+LSST+$H_0$) & $0.82$ & $2.1$ \\ \hline
Base + BAO + \Euclid & $0.80$ & $2.0$ \\ \hline
Base + BAO + DESI & $0.78$ & $1.9$ \\ \hline
Base + BAO + \Euclid + DESI & $0.75$ & $1.9$ \\ \hline
Data above + $y$-dist.~+ kSZ & $0.75$ & $1.7$ \\ \hline
\hline
\multicolumn{3}{c}{Curved background FLRW metric} \\ \hline
Data above & $0.82$ & $1.9$ \\
\hline
\hline
\end{tabular}}
\end{center}
\end{table}

We now present separately our results for the cases of the \lcdm and \lltb fiducial models of \Cref{tab:fiducials}. As said earlier, we use the \lcdm fiducial model to test how well future data can constrain deviations from the FLRW metric, while we use the \lltb fiducial models to see if future data can detect a violation of the Copernican principle.

\subsection{\lcdm mocks} \label{lambda-mocks}

\subsubsection{The Copernican principle in light of the forthcoming surveys}

\begin{figure*}[!tph]
\centering
\includegraphics[width = \textwidth]{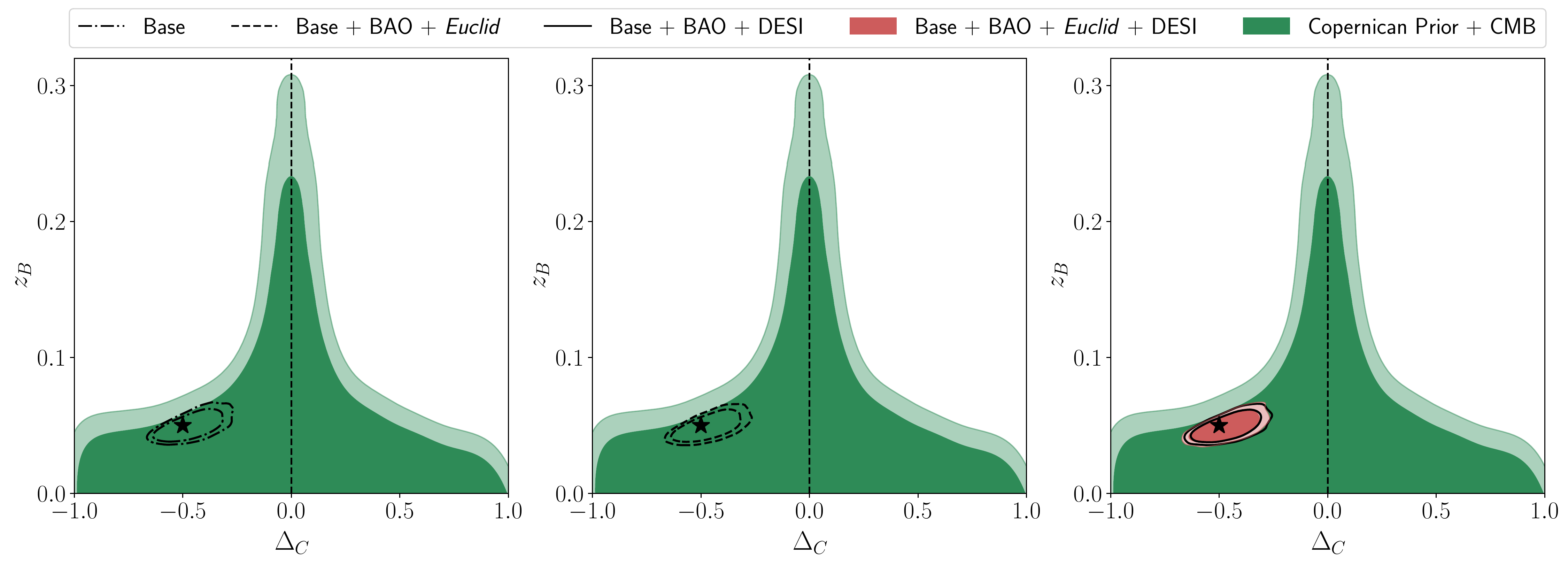}
\includegraphics[trim={0 0 0 1.65cm}, clip, width = \textwidth]{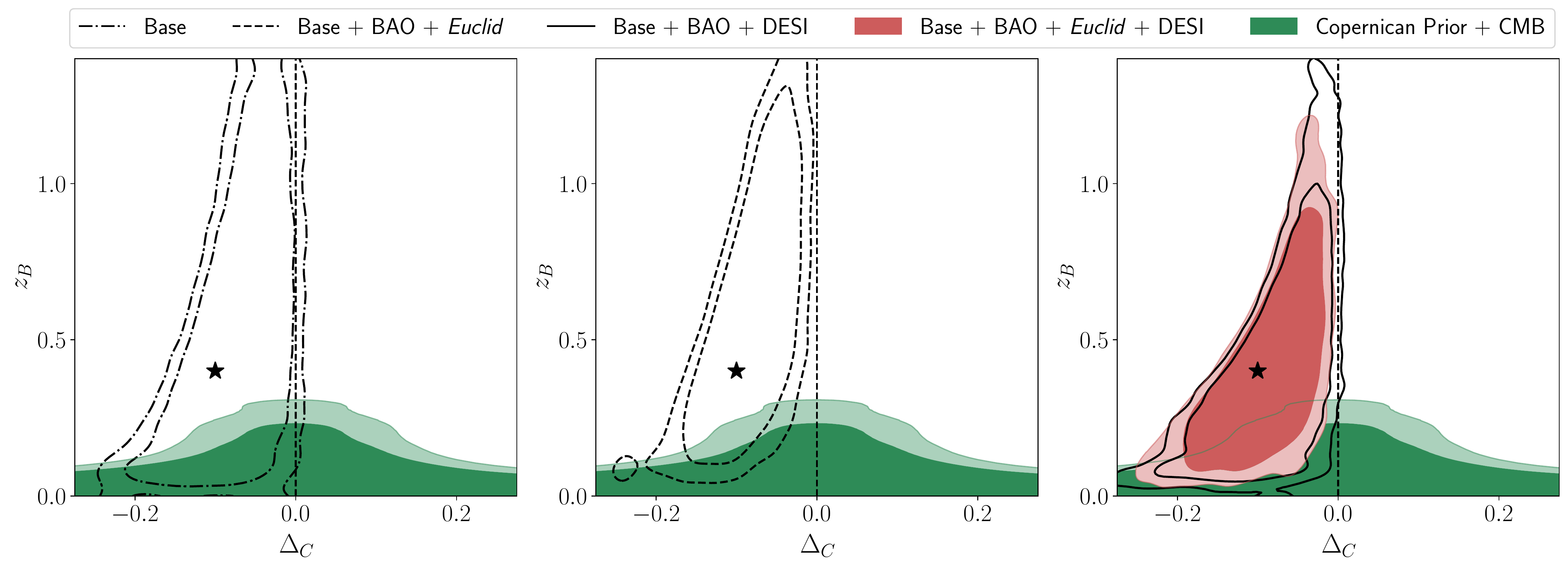}
\includegraphics[trim={0 0 0 1.65cm}, clip, width = \textwidth]{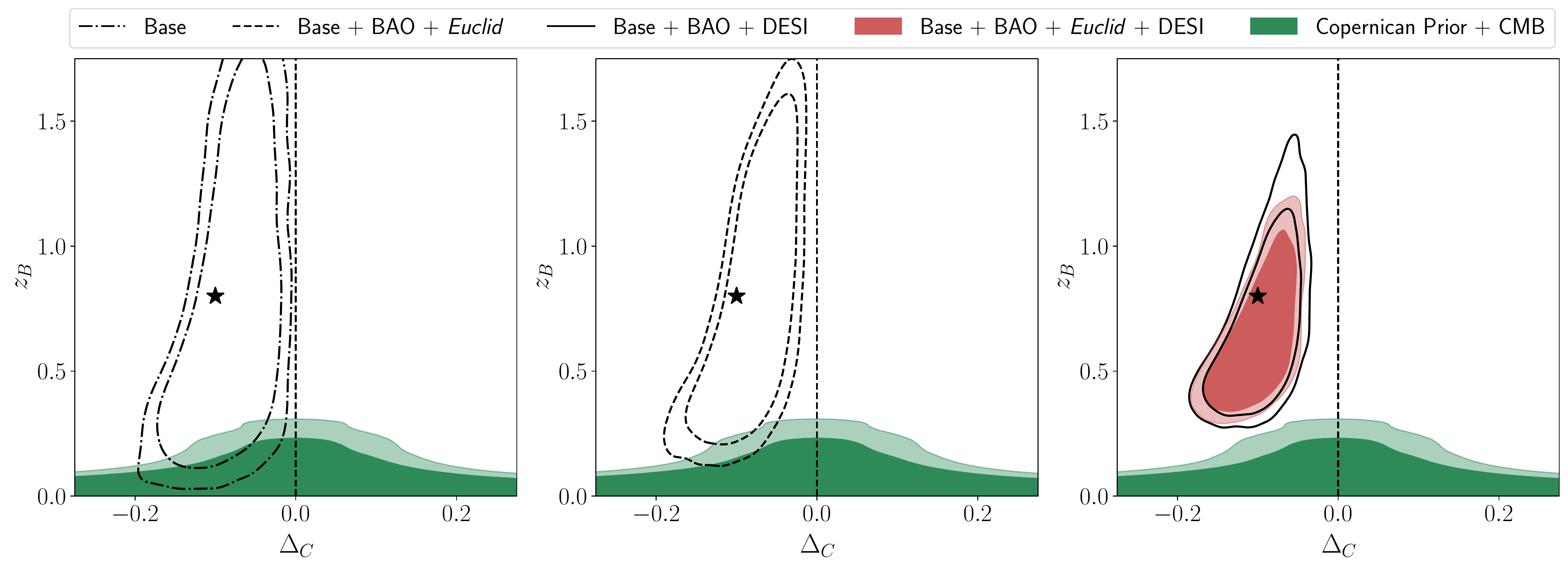}
\caption{ $95\%$ and $99\%$ confidence level constraints on the contrast at the centre, $\Delta_{\rm C}$, and the redshift of the boundary, $z_{\rm B}$, for the \lltb mock catalogues of \Cref{tab:fiducials} as compared to the constraints from the Copernican prior convolved with the CMB likelihood. The black star is placed at the fiducial values for the LTB parameters, i.e. $\Delta_{\rm C} = -0.5$ and $z_{\rm B} = 0.05$ (top row, \lltb~1), $\Delta_{\rm C} = -0.1$ and $z_{\rm B} = 0.4$ (middle row, \lltb~2), and $\Delta_{\rm C} = -0.1$ and $z_{\rm B} = 0.8$~(bottom row, \lltb~3).  We note that the $z_{\rm B}$ axis is not same for all figures.}
\label{fig:lltb1}
\end{figure*}

\begin{table*}[!tbh]
\begin{center}
\setlength{\tabcolsep}{15pt}
\renewcommand{\arraystretch}{1.35}
\caption{Percent improvement on constraints on radial inhomogeneity from next-generation surveys as compared to present-day constraints. \label{tab:curre_vs_fore}}
\small{\begin{tabular}{llcc}
\hline
\hline
\multirow{2}{*}{\parbox{6cm}{Observables considered in this analysis}} & \multirow{2}{*}{\parbox{5cm}{Present-day observables considered in \citet{Camarena:2021mjr}}} & \multicolumn{2}{c}{Percent improvement} \\
    &    & $0 \!\leq\! r^{\textrm{out}}_{\rm L}$  & $190\,\mathrm{Mpc} \!\leq\! r^{\textrm{out}}_{\rm L} $ \\ \hline
\hline
\multirow{3}{*}{Base} & Base C (CMB + Pantheon + $M_0$) & 29$\%$ & 28$\%$ \\
 & Base C + BAO + HZ & 26$\%$ & 28$\%$ \\
 & Base C + BAO + HZ + $y$-dist.~+ kSZ & 20$\%$ & 0$\%$ \\ \hline
\multirow{3}{*}{Base + BAO + \Euclid + DESI} & Base C & 35$\%$ & 34$\%$ \\
 & Base C + BAO + HZ & 32$\%$ & 34$\%$ \\
 & Base C + BAO + HZ + $y$-dist.~+ kSZ & 26$\%$ & 10$\%$ \\ \hline
\multirow{3}{*}{Base + BAO + \Euclid + DESI  + $y$-dist.~+ kSZ} & Base C & 35$\%$ & 41$\%$ \\
 & Base C + BAO + HZ & 32$\%$ & 41$\%$ \\
 & Base C + BAO + HZ + $y$-dist.~+ kSZ & 26$\%$ & 19$\%$ \\
\hline
\hline
\end{tabular}}
\end{center}
\end{table*}

In \Cref{fig:CP} we show the marginalised  constraints at the $95\%$ and $99\%$ confidence levels on the integrated mass contrast, $\Delta_{\rm L}$, and the comoving size, $r^\textrm{out}_{\rm L}$, for three different data combinations as compared to the constraints coming from the Copernican prior convolved with the CMB likelihoods.

The constraining power of future surveys on the radial inhomogeneity can be quantitatively compared to the expectation from the Copernican prior and CMB by comparing the ratio of the 95\% confidence regions in the parameter space (see \Cref{tab:Area}). Considering all scales, the ratio is always less than one, showing the capability of future surveys to rule out non-Copernican structures. However, at large scales, constraints provided by data still allow for non-Copernican mass density fluctuations since for $r^{\textrm{out}}_{\rm L} \ge 190\,{\rm Mpc}$ the ratio is approximately equal to two. We note that, for both cases, the combination Base + DESI + \Euclid provides constraints comparable to those obtained from the combination of all data, pointing out the important role that forthcoming LSS surveys will have to test the Copernican principle.

We also consider the case of non-zero background curvature, that is, $K_{\rm B} \neq 0$ in \Cref{eq:kr}. The result is shown in the last row of \Cref{tab:Area}. The inclusion of  background curvature  degrades the constraints by approximately $10\%$ compared to the flat case, still providing a competitive constraint on the non-Copernican parameters.

\subsubsection{Comparison with present-day constraints}

In order to quantify the role of future surveys in constraining inhomogeneity around us, we compare our constraints with the ones from current data only, as obtained in \citet{Camarena:2021mjr}.
Specifically, we compute the improvement on the observed area $A_\textrm{obs}$ considering the data combinations presented in~\Cref{tab:curre_vs_fore}. Our present analyses do not include a cosmic chronometer dataset as contributions of this kind of data are expected to be secondary as compared with SNe and BAOs \citep{Camarena:2021mjr}. We note that our previous implementation of such data did not include the full covariance matrix presented in \cite{Moresco:2020fbm}, revised and discussed in \cite{Moresco:2022phi}.

Our Base analysis shows an improvement upon the current constraints by more than $20\%$, when all scales are considered, and provides an improvement of $28\%$ when compared to the constraints from Base C and Base C + BAO + HZ at scales $r^{\textrm{out}}_{\rm L} \ge  190\,  \mathrm{Mpc}$, where HZ denotes the cosmic chronometers dataset used in \cite{Camarena:2021mjr}. It is interesting to note that our forecast Base analysis provides constraints comparable to those obtained with all the latest cosmological data available, Base C + BAO + HZ + $y$-dist + kSZ case, showing the importance of forthcoming SN surveys and $1\%$ prior on the Hubble constant.

On the other hand, LSS surveys will play an important role in testing the Copernican principle. As shown in~\Cref{tab:curre_vs_fore}, future measurements from \Euclid and DESI will sharpen the current constraints of Base C by approximately $35\%$, both at $0 \!\leq\! r^{\textrm{out}}_{\rm L}$ and $190\, \mathrm{Mpc} \!\leq\! r^{\textrm{out}}_{\rm L}$. The inclusion of \Euclid and DESI will also tighten the parameter space by more than $30\%$ compared to the combination Base C + BAO + HZ. When compared to the combination Base C + BAO + HZ + $y$-dist. + kSZ, our analysis with the forthcoming \Euclid and DESI data shows an improvement of $26\%$ for $0 \!\leq\! r^{\textrm{out}}_{\rm L}$ and $10\%$ for $190\, \mathrm{Mpc} \!\leq\! r^{\textrm{out}}_{\rm L}$.

Finally, the combination of all data considered here will  tighten our current constraints, leading to improvements up to $41\%$ for scales at $190\, \mathrm{Mpc} \!\leq\! r^{\textrm{out}}_{\rm L}$ and $35\%$ for $0 \!\leq\! r^{\textrm{out}}_{\rm L}$ (see \Cref{tab:curre_vs_fore}).

\subsection{\lltb mocks}

In \Cref{fig:lltb1} we show the marginalised  constraints at the $95\%$ and $99\%$ confidence levels on  $\Delta_{\rm C}$ and $z_{\rm B}$, for the three \lltb fiducial cosmologies, as compared to the constraints coming from the Copernican prior and CMB observations.

From the analysis relative to \lltb~1 (top row), we can see that future data will be able to probe the local structure. This means that the effect of the cosmic variance on the position of the observer will be reduced thanks to the forthcoming surveys.

On the other hand, from the analysis relative to \lltb~2 (middle row) and~3 (bottom row), we see that inhomogeneities that are large, but relatively shallow, can be detected with high significance thanks to future data. More precisely, one can note that our analyses exclude the FLRW case ($\Delta_{\rm C} = 0$ and $z_{\rm B} =0$) by $\gtrsim 3 \sigma$ (pink contours). This stresses the important roles of the next-generation surveys in testing the Copernican principle.
\\

\section{Discussion \label{sec:discussion}}

\subsection{The role of large-scale structure data} \label{subsec:BAO}

We have seen from the results of Sect.~\ref{lambda-mocks} on the \lcdm mocks that future surveys, such as \Euclid, will grant a $\approx\,$30\% improvement on inhomogeneity around the observer. In particular, for scales greater than 190 Mpc, the combination of all data  will constrain inhomogeneity to only 1.7 times the area of the region allowed by standard cosmology.
Given the fact that \Euclid probes the redshift range $0.9<z<1.8$, one may wonder if the improvement due to \Euclid comes directly from better constraints on the shape of the angular diameter distance and Hubble rate or indirectly from better constraints on the cosmological parameters.

\begin{figure*}[!tph]
\centering
\includegraphics[ width = \textwidth]{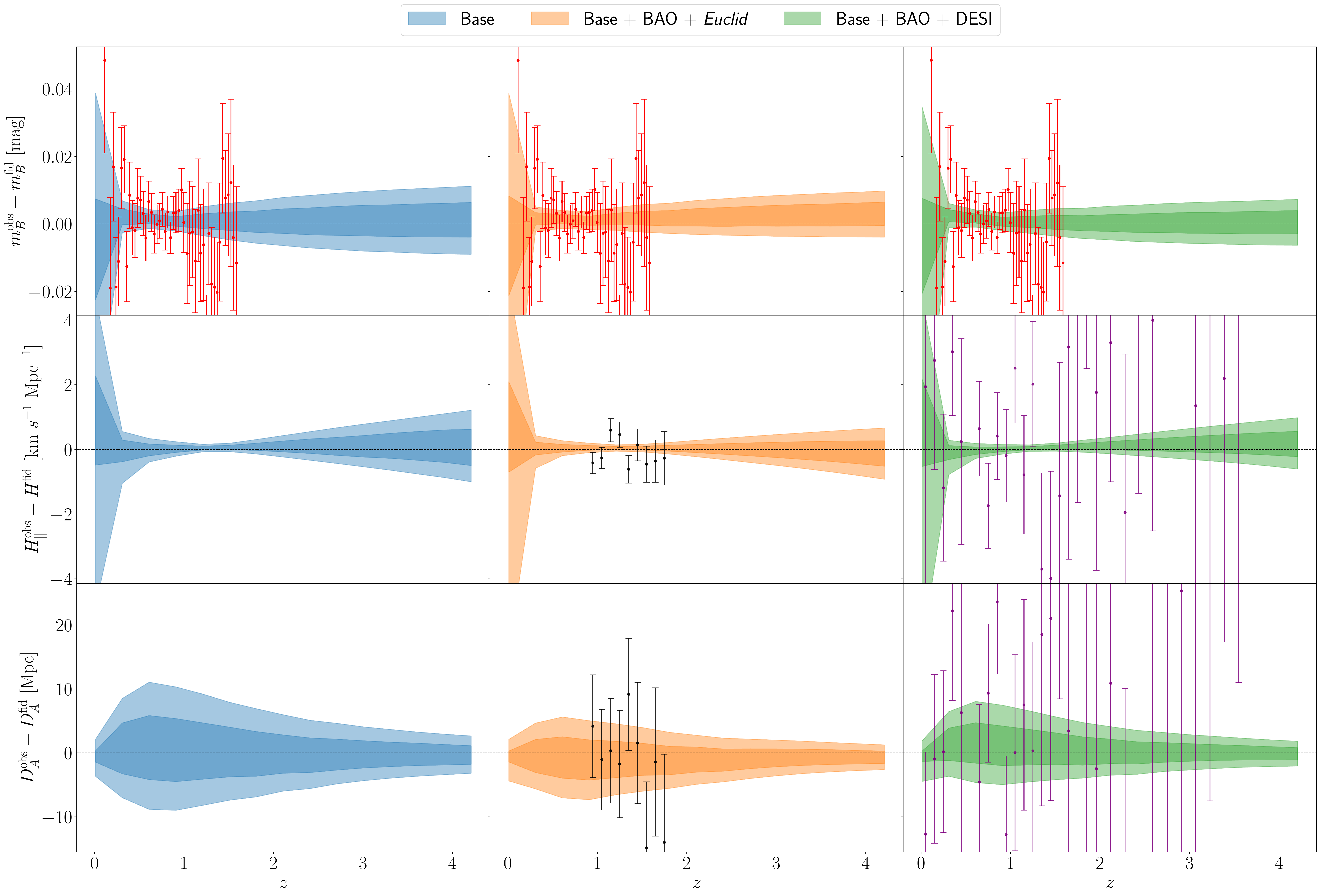}
\caption{Fluctuations in the apparent magnitude (top row), Hubble rate (middle row), and angular diameter distance (bottom row) for the \lltb model as compared to the fiducial \lcdm one. The 68\% and 95\% bands are obtained by evaluating the relevant functions at every point of the chains. The red points show LSST and \Euclid DESIRE SN data, the black ones \Euclid data, and the purple ones DESI data. See Sect.~\ref{subsec:BAO}.}
\label{fig:extra}
\end{figure*}

In order to answer the previous question we show in \Cref{fig:extra} the fluctuations in the apparent magnitude, Hubble rate and angular diameter distance for the \lltb model as compared to the fiducial \lcdm one. The 68\% and 95\% bands are obtained by evaluating the relevant functions at every point of the chains. We compare three analyses: the Base one, Base with present BAO and \Euclid, and Base with present BAOs and DESI.
From this plot, it appears that the shape of the various functions does not change when adding \Euclid or DESI. In other words, these two surveys do not improve the constraints in specific redshift ranges but rather they help at tightening the overall uncertainties. From this we conclude that the improvement due to \Euclid comes mostly from better constraints on the cosmological parameters, although this works in synergy with  DESI and the other observables.

\subsection{Beyond the central observer \label{subsec:centre}}

As mentioned earlier, our aim is to test radial homogeneity around us, neglecting anisotropies. We then placed the observer at the centre of the spherical over/underdensity.
However, in an inhomogeneous universe beyond FLRW, neglecting anisotropies could not be justified because anisotropies may affect observables as much as radial inhomogeneities.
In other words, the modelling adopted in this work implies a spherically symmetric inhomogeneity and a fine-tuning of the observer's position.

From the results of Sect.~\ref{lambda-mocks} on the \lcdm mocks we see, a posteriori, that large structures with shallow contrasts are allowed by future data. If, for example, we consider a contrast of $\delta =- 0.1$, the corresponding change in the Hubble rate is approximately $\delta H_0/H_0 = - f(\Omega_m)\delta/3  \simeq 0.017$, where $f\simeq 0.5$ is the present-day growth rate for the concordance $\Lambda$CDM model. The CMB dipole, if the observer were at, for example,~a distance $d_{\rm obs}=300$ Mpc from the centre, using $ v = \Delta H \, d_{\rm obs}$, is then
\begin{align}
\beta  = {v \over c} \simeq   1.2 \times 10^{-3} \,,
\end{align}
which is basically the observed CMB dipole \citep[][]{Planck:2018nkj}.
As the structures that we consider in this work extend to, at most, 1000 Mpc (see \Cref{fig:CP}), the required fine-tuning has a chance of less than 1 in 40. In other words, the fine-tuning required to satisfy the CMB dipole is rather mild and therefore the motivation for considering an off-centre observer is to provide a better description of possibly anisotropic data, rather than to relieve the fine-tuning of the observer’s position.

It is worth mentioning that the fine-tuning is instead very severe when considering void models as alternatives to dark energy, a possibility that was not explored here and not favoured by data (see \citealt[][]{Marra:2022ixf}). Indeed, in this case the underdensity has a radius of $\approx\,$3 Gpc and $\delta H_0/H_0 \approx\, 0.2$ so that the observer has to be within $\approx\,$30 Mpc from the centre, giving rise to a fine-tuning of one in a million \citep[][]{Marra:2011ct}.
We note, however, as pointed out in \citet{Garcia-Bellido:2008sdt}, that it is possible to alleviate this improbability by displacing the observer and then making them move towards the centre. For distances of a few hundred Mpc and velocities of a few thousand km\,s$^{-1}$, the effect is indistinguishable from the observed CMB dipole. In a way, one exchanges an improbability in location for an improbability in the direction of motion. The overall effect is to reduce the coincidence to a few parts in a thousand.

\section{Conclusions \label{sec:conclusions}}
Testing fundamental assumptions of cosmology is a crucial step towards improving our understanding of the Universe and firmly establishing the foundations of the standard cosmological paradigm. In this work we have tested the Copernican principle by placing constraints on the \lltb model using current and forecast data products. Specifically, we  focused on the capability of \Euclid to test the Copernican principle in conjunction with data from current and forthcoming surveys, such as SH0ES, DESI, and LSST.

In particular, we compared constraints on the \lltb model coming from the forecast and current data against constraints drawn from the  Copernican prior---the statistical counterpart of the Copernican principle. This comparison allowed us to quantify how well we can constrain deviations from the Copernican principle.

We have considered two types of fiducial models: the standard \lcdm model and the inhomogeneous \lltb model. By analysing the latter we aimed to determine if  next-generation surveys will be able to detect deviations from the Copernican principle, while our analysis of \lcdm data aimed to investigate if forthcoming data can successfully test the Copernican principle.

We have found that the inclusion of data from \Euclid , and other future surveys, will improve the current constraints on the Copernican principle by up to $40\%$. This improvement will be especially important at scales $r_{\rm B} \geq 190$ Mpc, where the inclusion of \Euclid, and other forthcoming surveys, will reduce the constrained area of the space parameters by a factor of $< 2$ as compared with the area allowed by the Copernican prior. Furthermore, we find that using the forthcoming \Euclid data, and data from other future surveys, we will be able to detect inhomogeneous deviations of the FLRW metric, including gigaparsec-scale inhomogeneities of contrast $-0.1$. Our analyses show that, given the precision of \Euclid and other forthcoming surveys, a detection of this kind would allow us to rule out the FLRW space-time ($\Delta_{\rm C} = 0$ and $z_{\rm B} =0$) by~$\gtrsim 3 \sigma$.

Our results rely on the assumption of a particular curvature profile, and, as shown in \Cref{app:kr}, constraints could be weakened by up to a factor of $\sim 2$ under the assumption of a more general profile. This drawback in our analysis, produced by the choice of a particular curvature profile, could be overcome by introducing data-driven methods that allow us to reconstruct the local distribution of matter in a more robust way. We will implement approaches of this sort in future research.

In summary, this work highlights the importance of synergies between \Euclid and external probes in testing the Copernican principle, which is one of the fundamental assumptions of the standard cosmological paradigm.

\begin{acknowledgements}
DC thanks CAPES for financial support. VM thanks CNPq and FAPES for partial financial support. This project has received funding from the European Union’s Horizon 2020 research and innovation programme under the Marie Skłodowska-Curie grant agreement No 888258. JGB, MM and SN acknowledge support from the research project  PGC2018-094773-B-C32, and the Spanish Research Agency (Agencia
Estatal de Investigaci\'on) through the Grant IFT Centro de Excelencia Severo
Ochoa No CEX2020-001007-S, funded by MCIN/AEI/10.13039/501100011033. LL was supported by a Swiss National Science Foundation Professorship grant (Nos.~170547 \& 202671). The work of CJAPM work was financed by FEDER---Fundo Europeu de Desenvolvimento Regional funds through the COMPETE 2020---Operational Programme for Competitiveness and Internationalisation (POCI), and by Portuguese funds through FCT - Funda\c c\~ao para a Ci\^encia e a Tecnologia in the framework of the project POCI-01-0145-FEDER-028987 and PTDC/FIS-AST/28987/2017. MM also received  support from ``la Caixa'' Foundation (ID 100010434), with fellowship code LCF/BQ/PI19/11690015. DS acknowledges financial support from the Fondecyt Regular project number 1200171. AdS acknowledges the support from the Fundação para a Ciência e a Tecnologia (FCT) through the Investigador FCT Contract No. IF/01135/2015 and POCH/FSE (EC) and in the form of an exploratory project with the same reference. JPM and AdS acknowledge the support from FCT Projects with references EXPL/FIS-AST/1368/2021, PTDC/FIS-AST/0054/2021, UIDB/04434/2020, UIDP/04434/2020, CERN/FIS-PAR/0037/2019, PTDC/FIS-OUT/29048/2017. 
This work made use of the CHE cluster, managed and funded by COSMO/CBPF/MCTI, with financial support from FINEP and FAPERJ, and operating at the Javier Magnin Computing Center/CBPF. 
This work also made use of the Virgo Cluster at Cosmo-ufes/UFES, which is funded by FAPES and administrated by Renan Alves de Oliveira.

\AckEC
\end{acknowledgements}

\bibliographystyle{aa} 
\bibliography{references} 

\clearpage
\begin{appendix} 

\section{The curvature profile \label{app:kr}}

The analyses presented in this paper rely on the assumption of the compensated profile introduced in \Cref{subsec:profile}, which was chosen in order to ensure that the \lcdm background is recovered at $r \ge r_{\rm B}$, a crucial feature in order to confront CMB data consistently using an effective FLRW model.

\begin{figure}[h!]
\centering
\includegraphics[width = \columnwidth]{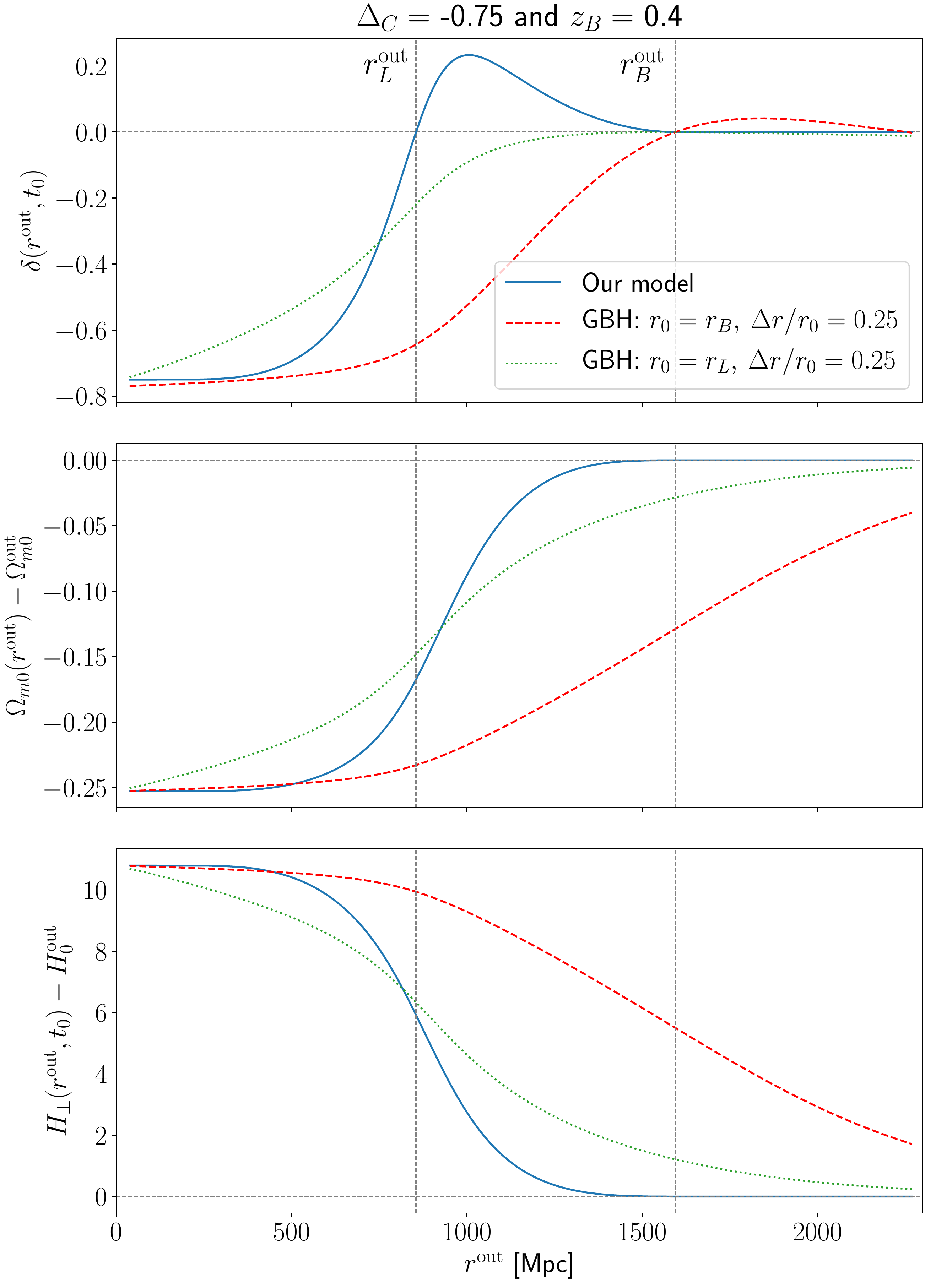}
\caption{Comparison between our model and the GBH parametrization. Top: Density contrast of matter today, $\delta (r,t_0)$, as a function of the FLRW comoving coordinate. Our compensated model (solid blue line) satisfies $\delta (r_{\rm L}^{\mathrm{out}},t_0) =  \delta (r_{\rm B}^{\mathrm{out}},t_0) = 0$, while the \lcdm background is not exactly recovered for the GBH models (dashed red and dotted green lines). Middle: Deviations from the background density of matter, $\Omega^{\mathrm{out}}_{\rm m,0}$, as a function of the FLRW comoving coordinate. The GBH model with $r_0 = r_{\rm B}$ (red dashed line) largely deviates from $\Omega^{\mathrm{out}}_{\rm m,0}$ at $r^{\mathrm{out}} = r_{\rm B}^{\mathrm{out}}$. On the other hand, the choice of $r_0 = r_{\rm L}$ for the GBH leads to deviations of approximately $-0.01$ from the background at $r^{\mathrm{out}} = r_{\rm B}^{\mathrm{out}}$. The model assumed in this work fully recovers $\Omega^{\mathrm{out}}_{\rm m,0}$ at any $r^{\mathrm{out}} \geq r_{\rm B}^{\mathrm{out}}$. Bottom:
Deviations from the background Hubble expansion as a function of the FLRW comoving coordinate. GBH models (dashed red and dotted green lines) do not perfectly match the \lcdm background expansion history at $r^{\mathrm{out}} = r_{\mathrm{B}}^{\mathrm{out}}$. \label{fig:GBH}}
\end{figure}

Here, we compare our model to the Garcia-Bellido and Haugboelle (GBH) model \citep{Garcia-Bellido:2008vdn}, which parametrises the LTB metric by imposing
\begin{align*}
\Omega_{\rm m,0}(r) & = \Omega^{\mathrm{out}}_{\rm m,0}  + \left( \Omega^{\mathrm{in}}_{\rm m,0} - \Omega^{\mathrm{out}}_{\rm m,0}\right) \left\lbrace \frac{1 - \tanh \left[ (r - r_0)/2 \Delta r \right] }{1 + \tanh \left[ r_0/2 \Delta r \right] } \right\rbrace\,, \\
H_0(r) & = \Hperpo^{\mathrm{out}} + \left( \Hperpo^{\mathrm{in}} -\Hperpo^{\mathrm{out}} \right) \left\lbrace \frac{1 - \tanh \left[ (r - r_0)/2 \Delta r \right] }{1 + \tanh \left[ r_0/2 \Delta r \right] } \right\rbrace  \,,
\end{align*}
where $r_0$ is the size of the void, $\Delta r$ the transition scale, $\Omega^{\mathrm{in}}_{\rm m,0} \equiv \Omega_{\rm m,0}(r=0)$, and $\Hperpo^{\mathrm{in}} \equiv \Hperpo (r=0)$.
\Cref{fig:GBH} shows the differences between our model (solid blue line) and the GBH model, where we have adopted $r_0 = r_{\rm L}$ (dotted green line) and $r_0  = r_{\rm B}$ (dashed red line).
When the size of the GBH inhomogeneity is fixed to $r_{\rm B}$, a scale greater that $r_{\rm B}$ is needed to recover the \lcdm background. On the other hand, if one assumes $r_0 = r_{\rm L}$, the GBH model tends to the \lcdm background at $r = r_{\rm B}$. In contrast, our model perfectly matches the \lcdm background at any scale $r \geq r_{\rm B}$. The compensating behaviour of our model is particularly notable in the top panel of the \Cref{fig:GBH}, where we note that $\delta (r_{\rm L}^{\mathrm{out}},t_0) =  \delta (r_{\rm B}^{\mathrm{out}},t_0) = 0$ for our model while the GBH models does not satisfy $\delta (r^{\mathrm{out}},t_0) = 0$ for all $r^{\mathrm{out}} \geq r_{\rm B}^{\mathrm{out}}$.

Furthermore, to investigate the dependence of our results on the chosen profile, we performed an additional analysis using the following generalisation of \Cref{eq:kr}:
\begin{align}
P_{3}(x,\alpha)=
\begin{cases}
1 &  \mbox{ for } 0 \le x < \alpha \\
1 - \exp \big[-\frac{1 - \alpha}{x - \alpha}(1- \frac{x - \alpha}{1 - \alpha})^3] & \mbox{ for }  \alpha  \le x < 1\\
0 & \mbox{ for } 1 \leq x 
\end{cases} \,, 
\end{align}
with $0 < \alpha < 1$. This new parameter will modify the smoothness of the transition between the inner and compensating region, leading to sharpened profiles when $\alpha$ approximates to $1$.
Results from this extra analysis, which is performed using the combination Base + BAO + \Euclid + DESI  + $y$-dist.~+ kSZ from the \lcdm fiducial, shows that the inclusion of the $\alpha$ parameter weakens the constraints on $\Delta_{\rm C}$ and $r_{\rm L}^{\mathrm{out}}$ by a factor of two, compared to the results from \Cref{eq:kr}. 

\section{Re-scaling datasets \label{app:mock_LLTB}}

Covariance matrices are fundamental pieces of forecast analyses. However, their production  for  forthcoming surveys is an open issue when non-standard cosmologies are considered \citep{Harnois-Deraps:2019rsd,DES:2020ypx,Ferreira:2021dmb}. This complicates the construction of forecast data for \lltb cosmologies. \citet{Euclid:2021frk} has overcome this issue by neglecting the error due to the non-standard cosmology. Here, we apply a re-scaling method to convert the \lcdm forecast data and its covariance matrices into \lltb catalogues.

Consider a given dataset, with $x_i$ being the observed quantity, $z_i$ the corresponding redshift, and $C_{ij}$ the covariance matrix. This dataset can be re-scaled to agree with a particular model via the following steps.

First, we define $R_{ij} =C_{ij}/x_i x_j$, a new matrix that contains the relative uncertainties and correlations from the original covariance matrix. Second, we compute with the theoretical prediction of the new model the fiducial values at the relevant redshifts, such that $x^{\mathrm{f}}_i \equiv x^{\mathrm{fid}}(z_i)$. Third, using the above defined quantities, we compute the new correlation matrix as $\tilde{C}_{ij} = x^{\mathrm{f}}_i  x^{\mathrm{f}}_j\: R_{ij}$.
Finally, we then draw a random realisation, $\tilde{x}_i$, of the multivariate-normal distribution $\mathcal{N}(x^{\mathrm{f}}_i, \tilde{C}_{ij})$.

We note that this method assumes that relative error and correlations are not changed by a non-standard model. As discussed through this paper, the procedure above is also applied to re-scale real data according the fiducial models presented on \Cref{tab:fiducials}; this ensures that all data are consistently described by a particular fiducial model.

\section{The inhomogeneous Hubble constant \label{app:HL}}

The $\Lambda$LTB model features a profile function $H_0(r)$ that depends on the radial distance from the centre of the void, instead of a constant value like $H_0$ in the \lcdm model. Since there is not a preferable scale to set the rate of expansion of the Universe, the definition of $H_0$ remains arbitrary. To overcome this issue, we extend the FLRW  definitions and mimic the observational procedure to locally constrain the Hubble constant. Explicitly, we adopt the definition $H_0^{\rm L}$ for inhomogeneous cosmological models that was introduced in \citet{Camarena:2022iae}. This method, which is applied for every sample point of the parameter space, follows the following steps. 

First, we create a mock catalogue using the redshifts of Pantheon SNe at $ 0.023 < z < 0.15$ and the $\Lambda$LTB luminosity distances at the corresponding redshifts. Second, the mock data are fitted using an extension of the cosmographic expansion given by
\begin{align}
D_L (z) & = \frac{cz}{H^{\rm L}_0} \left[ 1 + \frac{(1-q^r_0)z}{2} \right] \,, \\
q^r_0(r) & =  \left[ \frac{\Omega_{\rm m}(r)}{2} - \Omega_\Lambda(r) \right] \left[ \frac{H_0(r)}{H^{\rm L}_0} \right]^2 \,,
\end{align}
where $q^r_0$ is the radial-dependent deceleration parameter.
Finally, the best-fit value of $H_0^{\rm L}$ is adopted  as the measured Hubble constant.

It is interesting to point out that this procedure mimics the standard cosmic distance ladder analysis of SNe that follow the  Hubble flow, while taking into account the effect of the inhomogeneity on the measurement of the Hubble constant. We note that other authors have previously proposed similar approximations \citep{Redlich:2014gga,Efstathiou:2021ocp}.

\end{appendix}

\end{document}